\documentclass{article}

\usepackage[all,knot]{xy}
\usepackage{latexsym,graphics,epsfig}
\usepackage{rotate}
\usepackage{amsfonts,amssymb,amsmath}

\title{Exotic Statistics for Strings \\
       in 4d $BF$ Theory}

\author{John C.\ Baez and Derek K.\ Wise
\\ Department of Mathematics
\\ University of California
\\ Riverside, CA 92521, USA
\\
\\ email: baez@math.ucr.edu,
    derek@math.ucr.edu
\\
\\
Alissa S. Crans
\\Department of Mathematics
\\Loyola Marymount University/The Ohio State University
\\
\\ email:  acrans@lmu.edu
}
\date{April 20, 2006}

\newcommand{\R}{{\mathbb R}}
\newcommand{\C}{{\mathbb C}}

\newcommand{\Z}{{\mathbb Z}}

\def\tr {{\rm tr\,}}

\newcommand{\maps}{\colon}

\def\stackto #1 { \, {\stackrel{#1}{\longrightarrow}}\, }
\def\stackTo #1 { {\stackrel{#1}{\Longrightarrow}} }

\def\iso{\cong}

\newcommand{\SO}{{\rm SO}}
\newcommand{\so}{\mathfrak{so}}

\newcommand{\ISO}{{\rm ISO}}

\newcommand{\SL}{{\rm SL}}
\newcommand{\Sl}{\mathfrak{sl}}

\newcommand{\U}{{\rm U}}

\newcommand{\Aut}{{\rm Aut}}

\newcommand{\g}{\mathfrak{g}}

\newcommand{\zap}{\rhd}
\newcommand{\paz}{\lhd}

\newcommand{\A}{\mathcal{A}}
\newcommand{\G}{\mathcal{G}}

\newcommand{\Diff}{\mathrm{Diff}}

\newcommand{\Mo}{\mathrm{Mo}}

\def\hol {{\rm hol}}

\newcounter{letter} \newcounter{numeral} \newcounter{Numeral}

\newenvironment{alphalist}{
\begin{list}{(\alph{letter})}{\usecounter{letter}}
}{\end{list}}

\newsavebox{\Blor}
\savebox{\Blor}{\includegraphics{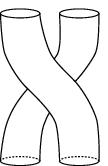}}
\newcommand{\blor}{\usebox{\Blor}}      

\newsavebox{\Brol}
\savebox{\Brol}{\includegraphics{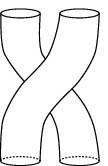}}
\newcommand{\brol}{\usebox{\Brol}}      

\newsavebox{\Bltrinv}
\savebox{\Bltrinv}{\includegraphics{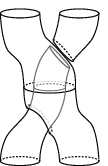}}
\newcommand{\bltrinv}{\usebox{\Bltrinv}}

\newsavebox{\Brtl}
\savebox{\Brtl}{\includegraphics{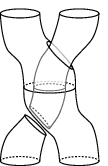}}
\newcommand{\brtl}{\usebox{\Brtl}}     

\newsavebox{\Brtlinv}
\savebox{\Brtlinv}{\includegraphics{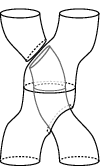}}
\newcommand{\brtlinv}{\usebox{\Brtlinv}}

\newsavebox{\Pureltr}
\savebox{\Pureltr}{\includegraphics{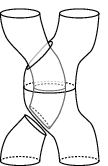}}
\newcommand{\pltr}{\usebox{\Pureltr}}

\newsavebox{\Pureltrinv}
\savebox{\Pureltrinv}{\includegraphics{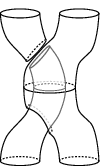}}
\newcommand{\pltrinv}{\usebox{\Pureltrinv}}

\newsavebox{\Purertl}
\savebox{\Purertl}{\includegraphics{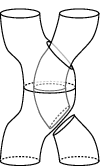}}
\newcommand{\prtl}{\usebox{\Purertl}}

\newsavebox{\Pipe}
\savebox{\Pipe}{\includegraphics{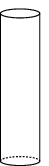}}
\newcommand{\pipe}{\usebox{\Pipe}}     

\newsavebox{\DoublePipe}
\savebox{\DoublePipe}{\includegraphics{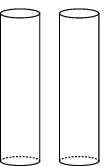}}
\newcommand{\dpipe}{\usebox{\DoublePipe}}    

\newcommand{\smalllabels}{\def\objectstyle{\scriptstyle}
                        \def\labelstyle{\scriptstyle}}

\newcommand{\twobytwo}[4]{\left(
     \begin{array}{cc}
        #1 & #2 \\
        #3 & #4
      \end{array}
      \right) }

\newcommand{\rot}[1]{\left(
     \begin{array}{cc}
        \cos {#1} & -\sin {#1} \\
        \sin{#1} & \cos {#1}
      \end{array}
      \right) }

\newtheorem{theorem}{Theorem}

\newcommand{\et}{\hspace{-0.08in}{\bf .}\hspace{0.1in}}
\def\qed {\hfill$\Box$\newline}
\newcommand\proof{\noindent\textbf{Proof.}\ }

\begin{document}
\maketitle

\begin{abstract}
After a review of exotic statistics for point particles in 3d $BF$ theory,
and especially 3d quantum gravity, we show that string-like defects 
in 4d $BF$ theory obey exotic statistics governed by the `loop braid 
group'.  This group has a set of generators that switch two strings 
just as one would normally switch point particles, but also a set of 
generators that switch two strings by passing one through the other.  
The first set generates a copy of the symmetric group, while the 
second generates a copy of the braid group.  Thanks to recent work 
of Xiao-Song Lin, we can give a presentation of the whole loop braid 
group, which turns out to be isomorphic to the `braid permutation 
group' of Fenn, Rim\'anyi and Rourke.  In the context 4d $BF$ theory 
this group naturally acts on the moduli space of flat $G$-bundles on 
the complement of a collection of unlinked unknotted circles in $\R^3$.  
When $G$ is unimodular, this gives a unitary representation of the 
loop braid group.  We also discuss `quandle field theory', in which 
the gauge group $G$ is replaced by a quandle.  
\end{abstract}

\newpage

\section{Introduction}

Physically speaking, the goal of this paper is to study the exotic 
statistics of loop-like defects in a 4-dimensional topological field 
theory called $BF$ theory.  We call these entities `closed strings' for
short, though they behave differently from the closed strings familiar 
in string theory: the relevant Lagrangian is different.  In fact, 
we postpone the study of their dynamics to another paper \cite{BaezPerez}.
The considerations of this paper are purely topological, and 
accessible---we hope---to mathematicians with only a passing interest
in physics.

Mathematically speaking, the point of this paper is to study some
representations of a higher-dimensional analogue of the braid group: 
the `loop braid group'.  Just as the braid group describes the 
topology of points moving in the plane, the loop braid group describes 
the topology of circles moving in $\R^3$.  In the body of this paper, 
we describe this group and certain representations of it coming from the 
moduli space of flat bundles on $\R^3$ with these circles removed.  But 
since everything we do has a more familiar analogue one dimension down, 
let us start by recalling that. 

{\boldmath
\subsubsection*{Exotic statistics in 3d $BF$ theory}
}

The behavior of a collection of identical particles
when they are exchanged goes by the name of `statistics'.
Traditionally, statistics was described using representations of
the symmetric group.  However, it is well known that in 3d spacetime,
`exotic' statistics are possible, in which the process of exchanging
identical particles is described by a representation of the braid group.
For example, exchanging two `abelian anyons' multiplies their wavefunction 
by a phase, which need not be $1$ as it is for bosons, nor $-1$ as for 
fermions.  This possibility has been investigated in experiments on the 
fractional quantum Hall effect \cite{CZG}.  Now researchers have begun 
the search for `nonabelian anyons', whose statistics are described by 
more complicated representations of the braid group \cite{BKS}.  Plans 
are already afoot to use these in quantum computers \cite{FKLW,Kitaev}.

Exotic statistics also arise naturally in the context of 3d
quantum gravity.  As we `turn on gravity', letting Newton's gravitational
constant $\kappa$ become nonzero, ordinary quantum field theory on 3d
Minkowski spacetime deforms into a theory where the Poincar\'e group
goes over to a quantum group called the $\kappa$-Poincar\'e group.
Moreover, if we begin with a field theory of bosons, their
statistics become exotic as we turn on gravity.
For a thorough treatment of these fascinating phenomena, see the papers
by Freidel and collaborators \cite{Freidel,FKS}, the paper by Krasnov
\cite{Krasnov}, and the many references therein.

In fact, the reason for exotic statistics in 3d quantum gravity is
very simple.  In 3d spacetime, Einstein's equations say that spacetime 
is {\it flat} except in regions where matter is present.  A point
particle at rest bends the nearby space into a cone.  This cone is 
flat everywhere except at its tip, where there is a deficit angle 
proportional to the particle's mass.  If we parallel 
transport a vector around the particle, it gets rotated by this 
angle $\theta$:
\[
\xy
(0,-1)*{\includegraphics{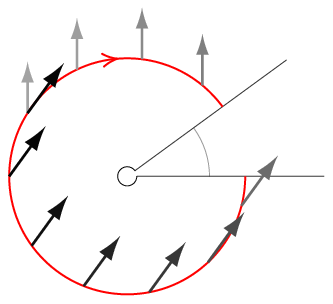}};
(-8.5,12)*{\theta};
(5,2)*{\theta};
\endxy
\]

More generally, if we have $n$ particles, space will be flat
except for conical singularities at $n$ points.
If we exchange these particles by moving them around the plane, 
they trace out a loop in the space of $n$-point subsets of the plane.  
Their energy-momenta will change in a way that depends on this loop---but 
only on the {\it homotopy class} of this loop, because they are being
parallel transported with respect to a flat connection.
A homotopy class of such loops is just an $n$-strand braid:
\[
\includegraphics{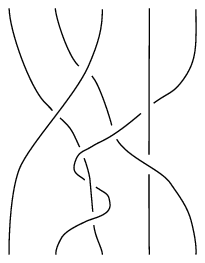}
\]
So, the group $B_n$ of $n$-strand braids acts on the Hilbert space of
states for $n$ identical particles.  In fact, this result holds
classically as well: we get an action of $B_n$ on the configuration
space for $n$ identical particles.

The above argument uses the fact that 3-dimensional gravity
(with vanishing cosmological constant) can be described by $BF$ theory
with the Lorentz group $\SO(2,1)$ as gauge group.  To understand this paper,
the reader only needs to know one thing about $BF$ theory:
{\it it involves a flat connection on space}.  For completeness,
however, we recall that $BF$ theory
in $n$-dimensional spacetime with gauge group $G$ involves two fields:
a connection $A$ and a $\g$-valued $(n-2)$-form $E$.
In the absence of matter, the Lagrangian is simply
\[    L = \frac{1}{\kappa} \tr(E \wedge F)  \]
Here $\kappa$ plays the role of Newton's constant in the case of 3d
gravity, and $F = dA + A \wedge A$ is the curvature of $A$.
The resulting equations of motion:
\[    F = 0, \qquad   dE + [A,E] = 0,   \]
imply that the connection $A$ is flat.

In 3d $BF$ theory, point particles can be included by considering
spacetimes with curves removed: we think of these as the particles'
worldlines.  Away from these worldlines the above equations still hold,
while along the worldlines $A$ becomes singular.  The holonomy around
a loop circling a worldline gives an element of the group $G$.
A collection of $n$ particles in the plane thus gives rise to
an $n$-tuple of elements of $G$.  For simplicity, consider the case
$n = 2$.  As we exchange two particles by rotating them around each
other counterclockwise, they trace out this braid:
\[
   \xy 0;<1cm,0cm>:
      \vcross~{(-.4,.75)}{(.4,.75)}{(-.4,-.75)}{(.4,-.75)};
    \endxy
\]
As we recall in Section \ref{3d},
this operation acts as the following map on $G^2$:
\begin{equation}
\label{braiding}
   (g_1,g_2) \mapsto (g_1 g_2 g_1^{-1}, g_1 )  .
\end{equation}
Applying this map twice does {\it not} give the identity, so
we do not obtain an action of the symmetric group on $G^2$, but
only an action of the braid group.  In other words,
the particles have exotic statistics!

In the case of 3d gravity, the singularity of the connection
along a particle's worldline reflects the fact that the particle's
mass creates a conical singularity in the metric.
The holonomy around the worldline, an element of $G = \SO(2,1)$,
describes the particle's {\it energy-momentum}.  This may
seem odd, since we are used to thinking of energy-momentum as a
vector in Minkowski spacetime.  However, in 3 dimensions Minkowski spacetime
is naturally isomorphic to the Lie algebra $\so(2,1)$, and we
can reinterpret Lie algebra elements as group elements via the map:
\[
\begin{array}{ccl}
          \so(2,1) &\to& \SO(2,1)   \\
              p    &\mapsto& \exp(\kappa p)  .
\end{array}
\]
So, we can encode the energy-momentum $p$ of a particle in the holonomy
$g = \exp(\kappa p)$ resulting from parallel transport around this
particle's worldline.

Thanks to the factor of $\kappa$ here, the group $\SO(2,1)$
effectively `flattens out' to $\so(2,1)$ in the $\kappa \to 0$ limit.
For example, multiplication in the group reduces to addition in the Lie
algebra plus small corrections:
\begin{equation}
\label{doubly_special}
   \exp(\kappa p_1) \exp(\kappa p_2) = \exp(\kappa (p_1 + p_2) +
\frac{\kappa^2}{2} [p_1,p_2]  + \cdots )
\end{equation}
This implies that in terms of $\so(2,1)$-valued energy-momenta, the
braiding in equation (\ref{braiding}) is given by
\[   (p_1, p_2) \mapsto (p_2 + \kappa [p_1,p_2] +\cdots \,, \, p_1 )
\]
So, the exotic statistics reduce to ordinary bosonic statistics
in the limit where Newton's constant goes to zero.  They also reduce
to bosonic statistics in the limit where the particles are at rest
relative to each other, since then $p_1$ and $p_2$ become
proportional and their commutator vanishes.

The corrections to the usual law for addition of energy-momenta
implicit in equation (\ref{doubly_special}) are interesting in 
themselves.  Like the exotic statistics, these corrections 
become negligible in the limit $\kappa \to 0$.  Under the name 
of `doubly special relativity', modified laws for adding 
energy-momentum have already been studied by many authors.
The paper by Freidel, Kowalski-Glikman and Smolin \cite{FKS}
gives a good account of doubly special relativity in the context
of 3d quantum gravity; their paper also explains more of the history 
of this subject.

\hskip 1em

\subsubsection*{Quandle field theory}

Besides exotic statistics and corrections to the usual rule for adding
energy-momenta, there is yet another surprising consequence
of the switch from vector-valued to group-valued energy-momentum as
we turn on gravity in 3d physics.  The classification of elementary
particles changes!

In ordinary quantum field theory on Minkowski
spacetime, the Lorentz group acts on the space of possible energy-momenta,
and the orbits of this action correspond to different types of
spin-zero particles.  When spacetime is 3-dimensional, the
space of energy-momenta is $\so(2,1)$, and the orbits look like this:
\[
\xy 0;<1cm,0cm>:
(0,0)*{\includegraphics{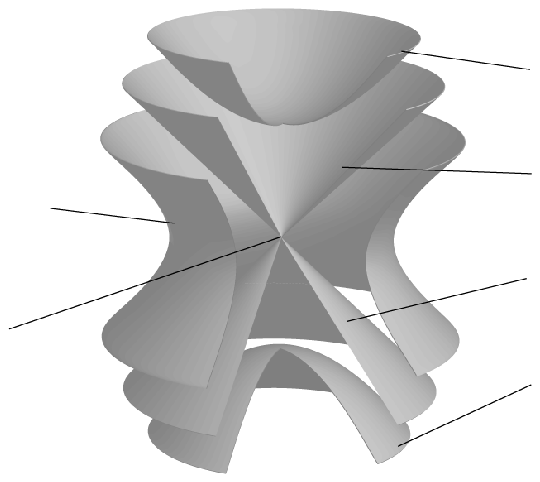}};
(4.6,1.8)*{\text{positive-energy tardyons}};
(4.6,-1.5)*{\text{negative-energy tardyons}};
(4.6,.7)*{\text{positive-energy luxons\phantom{xx}}};
(4.6,-.3)*{\text{negative-energy luxons\phantom{xx}}};
(-3.2,.4)*{\text{tachyons}};
(-3.3,-1.2)*{\text{particles of zero}};
(-3.3,-1.5)*{\text{energy-momentum}};
\endxy
\]
If we write the energy-momentum as $p = (E,p_x,p_y)$ and let
$p \cdot p = E^2 - p_x^2 - p_y^2$, we have six families
of orbits, corresponding to six types of spin-zero particles:

\begin{enumerate}
\item positive-energy tardyons of mass $m > 0$:
$\{p \cdot p = m^2, \; E > 0\}$,
\item negative-energy tardyons of mass $m > 0$:
$\{p \cdot p = m^2, \; E < 0\}$,
\item positive-energy luxons:
$\{p \cdot p = 0, \; E > 0\}$,
\item negative-energy luxons:
$\{p \cdot p = 0, \; E < 0\}$,
\item tachyons of mass $im$ for $m > 0$:
$\{p \cdot p = -m^2\}$,
\item particles of vanishing energy-momentum:
$\{p = 0 \}$.
\end{enumerate}
Given any orbit $Q \subseteq \so(2,1)$, the Hilbert space for a
single particle of type $Q$ is just $L^2(Q)$.

The same philosophy applies when we turn on gravity, but now the
space of energy-momenta is not the Lie algebra $\so(2,1)$ but the
Lorentz group itself.  This acts on itself by conjugation,
and the orbits are conjugacy classes.  Types of spin-zero particles
now correspond to conjugacy classes in the Lorentz group.
Near the identity these conjugacy classes look just like orbits in
the Lie algebra, so the classification of particles reduces to the
above one in the limit of small energy-momenta.
However, there are important differences, which show up for
large energy-momenta.

Most notably, under the map
\[                p \mapsto \exp(\kappa p) \]
the Lie algebra element $p = (E,0,0)$ is mapped to
a rotation by the angle $\kappa E$ in the $xy$ plane.   So, the
holonomy around a stationary particle of energy $E$ is a rotation
by the angle $\kappa E$.   This rotation does not change when we
add $2\pi/\kappa$ to the particle's energy.  Up to factors of order unity,
this quantity $2\pi/\kappa$ is just the {\it Planck energy}.  If we
call it the Planck energy, then masses in 3d quantum gravity
are defined only modulo the Planck mass.

This `periodicity of mass' affects the
classification of tardyons---that is, the most familiar sort of particles,
those with timelike energy-momentum.   Instead of positive-energy tardyons
of arbitrary mass $m > 0$ and negative-energy tardyons of arbitrary
mass $m > 0$, we just have tardyons of
arbitrary mass $m \in \R/{2\pi\over \kappa}\Z$.

More generally, for any Lie group $G$, the various allowed
types of spin-zero particles in 3d $BF$ theory with gauge group $G$
correspond to conjugacy classes $Q \subseteq G$.  Any conjugacy
class is closed under the operations
\[             g \zap h = ghg^{-1}   , \qquad h \paz g = g^{-1}hg , \]
and these operations satisfy equations making $Q$ into an algebraic structure
called a `quandle' \cite{J}, whose definition we recall in Section
\ref{quandle}.  The Hilbert space for a single particle
of type $Q$ is just $L^2(Q)$, defined using a measure on $Q$ that
is invariant under these operations.  In an easy generalization of
3d $BF$ theory, we can study the exotic statistics of `particles of
type $Q$' for any quandle $Q$ equipped with an invariant measure.
This takes advantage of the well-known relation between quandles and
the braid group \cite{FR}.

{\boldmath
\subsubsection*{Exotic statistics in 4d $BF$ theory}
}
It would be wonderful to generalize all the above results to
4d gravity, but for now all we can handle is a simpler theory:
4d $BF$ theory.   This may eventually be relevant to gravity,
since one can describe general relativity in 4 dimensions either
as the result of {\it constraining} 4d $BF$ theory with a certain
gauge group, or {\it perturbing around} 4d $BF$ theory with some
other gauge group.  The first approach goes back to
Plebanski \cite{Plebanski}, and it underlies a great deal of
work on spin foam models of quantum gravity \cite{Baez,Oriti,Perez},
especially the Barrett--Crane model.   The second approach
goes back to MacDowell and Mansouri \cite{MM}, and has recently
been explored by Freidel and Starodubtsev \cite{FreidelStarodubtsev}.
However, we do not dwell on these possible applications here.
They only focus our attention towards certain choices of gauge group:
\[
\begin{array}{rl}
\textrm{Plebanski gravity:} & \quad\; G = \SO(3,1)
\\
\\ \textrm{MacDowell--Mansouri gravity:}  &
\left\{
\begin{array}{ll}
G = \SO(4,1) & \Lambda > 0 \\
G = \SO(3,2) & \Lambda < 0
\end{array}
\right.
\end{array}
\]

Our idea is simply to increase the dimension of everything in the previous
section by 1.  Thus, we consider $BF$ theory on a 4-dimensional
spacetime with the worldsheets of several `closed strings' removed.
We focus on the case where the manifold representing space is $\R^3 -
\Sigma$, where $\Sigma$ is an `$n$-component unlink': a collection
of $n$ unknotted unlinked circles.   A flat connection on $\R^3 - \Sigma$
gives us a group element for each circle, namely the holonomy of
some standard loop going around this circle:
\[
\xy
(0,0)*{\includegraphics{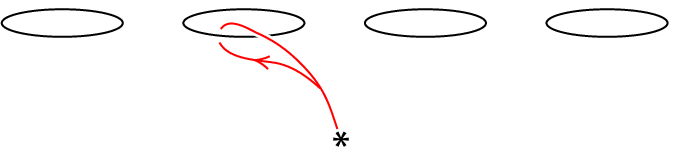}};
\endxy
\]

\noindent
So, just as before, we obtain $n$-tuples of elements of $G$.
Moreover, any way to exchange the circles in $\Sigma$ gives
a map from $G^n$ to itself.

It is often said that exotic statistics are only possible when
space has dimension 2 or less.  However, this folklore only 
applies to point particles.  As pointed out by Balanchandran and
others \cite{ABKS,Balachandran,Niemi,Surya,Szabo}, exotic statistics 
are possible for closed strings in 3-dimensional space, since 
there are topologically nontrivial ways to exchange unknotted unlinked 
circles in $\R^3$.  The statistics of such theories are governed 
not by the braid group $B_n$, but by a larger group: the `loop 
braid group' $LB_n$.   

Using recent work of Lin \cite{Lin}, we show that this group is isomorphic to
the `braid permutation group' of Fenn, Rim\'anyi and Rourke \cite{FRR}.
This is an apt name, because $LB_n$ has a presentation with
generators $s_i$ that describe two strings trading places without
passing through each other, just as if they were point particles:
\[
\xy  0;<1cm,0cm>:
(0,0)*{\brol};
(1,0)*{=};
(2,0)*{\blor};
\endxy
\]
but also generators $\sigma_i$ that describe one string passing through
another:
\[
\xy  0;<1cm,0cm>:
(0,0)*{\brtl};
(1,0)*{\neq};
(2,0)*{\bltrinv};
\endxy
\]
So, this group is a kind of `hybrid' of the symmetric
group and the braid group.  Indeed, the elements $s_i$ generate a
copy of the symmetric group $S_n$ in $LB_n$, while the elements $\sigma_i$
generate a copy of the braid group $B_n$.

In a one-dimensional unitary representation of the loop braid
group, the permutation generators $s_i$ all act as
$\pm 1$, while the braid generators $\sigma_i$ all act
as an arbitrary phase $q \in \U(1)$.   We could call particles
that transform in this way `abelian bose-anyons' and `abelian
fermi-anyons', respectively.  They act like bosons or fermions
when we switch them using the generators $s_i$, but like
abelian anyons when we switch them using the generators $\sigma_i$.

$BF$ theory gives us more interesting unitary representations
of the loop braid group: whenever the group $G$ is unimodular,
we obtain a unitary representation of $LB_n$ on $L^2(G^n)$.
All the groups listed above are unimodular, so we get an
interesting variety of exotic statistics for closed strings in
4d $BF$ theory.

We can also restrict attention to a specific conjugacy class
$Q \subseteq G$ and get a unitary representation of the loop
braid group on $L^2(Q^n)$, as long as $Q$ is equipped with a
measure invariant under conjugation.  As already mentioned,
in the case of 3d gravity a choice of conjugacy class in $G = \SO(2,1)$
essentially amounts to choosing a specific {\it mass} for our
point particles, which is a very natural thing to do.  In the
case of 4d $BF$ theory with $G = \SO(3,1)$, choosing a conjugacy
class essentially amounts to choosing a specific {\it mass density}
for our closed strings.

\hskip 1em

\section{The Loop Braid Group}
\label{loop braid group}

The loop braid group $LB_n$ consists of all ways a collection
of oriented, unknotted, unlinked circles can move around in $\R^3$
and come back to their original positions, perhaps trading places.
More precisely, it consists of `isotopy classes' of such motions.
This group thus plays the same role in describing the interchange
of closed strings in $\R^3$ that the symmetric group $S_n$ plays for
point particles in $\R^3$, and the braid group plays for
point particles in $\R^2$.  In this section we use the work of
Lin \cite{Lin} to obtain two presentations of the loop braid group.
First, however, we explain the sense in which the loop braid
group, the symmetric group and the braid group are all examples of
`motion groups'.

The general idea of a `motion group' goes back at least to Dahm's
1962 thesis \cite{Dahm}, which unfortunately was never published.
In the 1970's and 80's, some papers by Wattenberg
\cite{Wattenberg} and Goldsmith \cite{Goldsmith,Goldsmith2}
clarified and expanded on Dahm's work.  More recently, McCool
\cite{McCool} and Rubinsztein \cite{Rubinsztein} have studied
the motion group for unknotted and unlinked circles in $\R^3$.
Surya has also given a description of the loop braid group as
an iterated semidirect product \cite{Surya}.  Much of this work
considers the motion of unoriented circles.  Since we use oriented 
circles, we obtain a smaller motion group, which lacks the 
`circle-flipping' operations that reverse orientations.

Quite generally, suppose that $S$ is a smooth oriented manifold and
$\Sigma \subseteq S$ is a smooth oriented submanifold.  Let
$\Diff(S)$ be the group of orientation-preserving diffeomorphisms of
$S$.  Let $\Diff(S,\Sigma)$ be the subgroup of $\Diff(S)$
maps that restrict to give orientation-preserving diffeomorphisms
of $\Sigma$.

We define a {\bf motion} of $\Sigma$ in $S$ to be a smooth map
$f \maps [0,1] \times S \to S$, which we write as
$f_t \maps S \to S$ ($t \in [0,1]$), with the following properties:
\begin{itemize}
\item for all $t$, $f_t$ lies in $\Diff(S)$;
\item for all $t$ sufficiently close to $0$, $f_t$ is the identity;
\item for all $t$ sufficiently close to $1$,
$f_t$ is independent of $t$ and lies in $\Diff(S,\Sigma)$.
\end{itemize}
Intuitively, a motion is a way of moving $\Sigma$ through $S$ so
that it comes back to itself---not pointwise, but as a set---at $t = 1$.
This suggests that one can `multiply' motions by doing one after the
other, and indeed this is true.  Given motions $f$ and $g$, one
can define a motion $f\cdot g$ called their {\bf product} as follows:
\[
(f \cdot g)_t = \left\{
\begin{array}{lcl}
f_{2t}             & {\rm for} & 0 \le t     \le \frac{1}{2}  \\
g_{2t-1} \circ f_1 & {\rm for} & \frac{1}{2} \le t \le 1
\end{array}
\right.
\]
Given a motion $f$ we can also define a motion called its
{\bf reverse}, denoted $\bar{f}$, by:
\[     \bar{f}_t = f_{1-t} \circ f_1^{-1}  .\]
We say two motions $f$ and $g$ are {\bf equivalent} if
$\bar{f} \cdot g$ is smoothly homotopic, as a path in
$\Diff(S)$ with fixed endpoints, to a path that lies
entirely in $\Diff(S,\Sigma)$.   One can check that this is
indeed an equivalence relation and that the operations of
product and reverse make equivalence classes of motions into
a group.  This is called the {\bf motion group} $\Mo(S,\Sigma)$.

Next we turn to examples:
\begin{itemize}
\item
When $\Sigma \subset \R^d$
is a collection of $n$ points and $d > 2$, $\Mo(\R^d,\Sigma)$ is
the symmetric group $S_n$.
\item
When $\Sigma \subset \R^2$
is a collection of $n$ points, $\Mo(\R^2,\Sigma)$ is the
braid group $B_n$.
\item
When $\Sigma \subset \R^3$
is a collection of $n$ unknotted and unlinked oriented circles, we
call $\Mo(\R^3,\Sigma)$ the {\bf loop braid group} $LB_n$.
\end{itemize}

We shall use the work of Lin \cite{Lin} to give two presentations
of $LB_n$.   First note that there is a homomorphism
\[         p \maps LB_n \to S_n  \]
which simply forgets the details of the braiding,
remembering only how the circles get permuted in the process.
The image of $p$ is all of $S_n$.  We call the kernel of $p$ the
{\bf pure loop braid group} $PLB_n$.

Suppose, just to be specific, that $\Sigma = \ell_1 \cup \cdots
\cup \ell_n$ where $\ell_1, \dots, \ell_n$ are disjoint unit circles
in the $xy$ plane, lined up from left to right with their
centers on the $x$ axis.
Lin proves that $PLB_n$ has a presentation with generators $\sigma_{ij}$
for $i,j \in \{1,\dots,n\}$ with $i \ne j$.  The generator $\sigma_{ij}$
describes a motion in which the $i$th circle floats up and over
the $j$th circle, shrinks slightly and passes down through the $j$th
circle, expands to its original size, and then moves straight back
to its starting position.  We draw this as follows:
\[
   \sigma_{ij} =
   \xy 0;<1cm,0cm>:
      (0,0)*{\pltr};
      (-.35,1)*{i};
      (.35,1)*{j};
    \endxy
\]
where for purely artistic reasons we let the $j$th circle move
a bit to the left in the process.

Here we are using a drawing style adapted
from Carter and Saito's work on surfaces in 4 dimensions \cite{CS}.
Crossings in a braid or knot are usually drawn with an
artificial `break' in one of the strands to indicate that it lies
under the other:
\[
\xy
\xoverv~{(-5,5)}{(5,5)}{(-5,-5)}{(5,-5)};
\endxy
\]
Similarly, Carter and Saito draw 3d projections of knotted surfaces
in 4 dimensions, indicating by a broken surface which one passes
`under' the other in the suppressed fourth dimension.
In our context, we take this suppressed dimension to be one of the spatial
dimensions, in order to make room for {\em time}, which we decree to flow
downward in all our diagrams.  The broken surfaces in $\sigma_{ij}$ indicate
whether one circle is above or below the other in the suppressed spatial
dimension, so that the following diagram and `movie' illustrate the same
process:
\[
\xy 0;<1cm,0cm>:
   (2,0)*{\includegraphics[height=4.8cm]{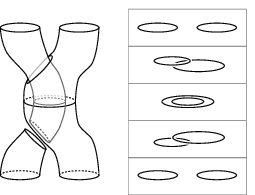}};
\endxy
\]
The inverse of $\sigma_{ij}$ is of course obtained by running the movie
backwards, which in diagrammatic notation becomes:
\[
\smalllabels
   {\sigma_{ij}}^{-1} =
   \xy 0;<1cm,0cm>:
      (0,0)*{\pltrinv};
      (-.3,1)*{i};
      (.3,1)*{j}
   \endxy
\]
One advantage of this drawing style is that it immediately suggests
Reidemeister--like moves for loop braids, such as this:
\[
 \xy
 (0,0)*{\includegraphics{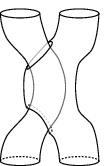}};
 \endxy
\quad
=
\quad
 \xy
 (0,0)*{\dpipe};
 \endxy
\quad
=
\quad
 \xy
 (0,0)*{\includegraphics{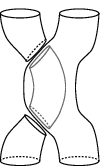}};
 \endxy
\]
We shall study the loop braid group algebraically, relying on such
diagrams for our intuition.

Given Lin's presentation of $PLB_n$, we can obtain a presentation
of $LB_n$ using the short exact sequence
\[       1 \to PLB_n \stackto{i} LB_n \stackto{p} S_n \to 1 . \]
First, note that there is a homomorphism
\[
j \maps S_n \to LB_n
\]
which takes a given permutation to what Lin calls a `permutation
path' in the motion group: a loop braid in which circles trade
places without any circle passing through another in a
topologically nontrivial way. For example, we can have them trade
places while remaining on the $xy$ plane. This map $j$ is
well-defined since all such permutation paths are homotopic.
Moreover, the composite $p\circ j \maps S_n \to S_n$ is the
identity homomorphism on $S_n$, so $j$ is a splitting of the short
exact sequence above.

Since $j$ is one-to-one, we may identify elements of
$S_n$ with their images in $LB_n$.  Since $PLB_n$ is a normal
subgroup, elements of $S_n$ act on $PLB_n$ via conjugation.
This allows us to define the semidirect product $S_n \ltimes PLB_n$,
and thanks to our split exact sequence, we get an isomorphism
\begin{align*}
    f \maps LB_n &\to S_n \ltimes PLB_n \\
       g &\mapsto (p(g),j(p(g))^{-1}g)
\end{align*}
with inverse
\begin{align*}
    f^{-1}\maps S_n \ltimes PLB_n &\to LB_n \\
                 (s,\sigma) &\mapsto s\sigma
\end{align*}

Writing the loop braid group as a semidirect product
in this way, we easily obtain a presentation for it:

\newpage 

\begin{theorem} \et \label{presentation1}
The loop braid group $LB_n$ has a
presentation with generators $s_i$ for $1 \le i \le n-1$ and
$\sigma_{ij}$ for $1 \le i,j \le n$ with $i \ne j$, together
with the following relations:
\begin{alphalist}

\item
the relations for the standard generators $s_i$ of $S_n$:
\begin{align}
\label{rel1}
s_i s_j &= s_j s_i                      &&
{\rm for \; } |i-j| > 1                 \\
\label{rel2}
s_i s_{i+1} s_i &= s_{i+1} s_i s_{i+1}  &&
{\rm for \; } 1 \le i \le n-2           \\
\label{rel3}
s_i^2 &= 1                              &&
{\rm for \; } 1 \le i \le n-1
\end{align}

\item
Lin's relations for the generators $\sigma_{ij}$ of $PLB_n$:
\begin{align}
\label{rel4}
\sigma_{ij}\sigma_{k\ell} &= \sigma_{k\ell}\sigma_{ij}  &&
{\rm  for \;} i,j,k,\ell {\rm  \; distinct}             \\
\label{rel5}
\sigma_{ik}\sigma_{jk} &= \sigma_{jk}\sigma_{ik}        &&
{\rm  for \;} i,j,k {\rm  \; distinct}                  \\
\label{rel6}
\sigma_{ij}\sigma_{kj}\sigma_{ik} &= \sigma_{ik}\sigma_{kj}\sigma_{ij}  &&
{\rm for \;} i,j,k {\rm  \; distinct}
\end{align}

\item
relations expressing the action of $S_n$ on $PLB_n$:
\begin{align}
\label{rel7}
 s_i \sigma_{i(i+1)} &=  \sigma_{(i+1)i} s_i            &&
{\rm for \; } 1 \le i \le n-1                          \\
\label{rel8}
s_k \sigma_{ij}  &=  \sigma_{ij} s_k                     &&
{\rm  for \;} i,j,k,k+1  {\rm  \; distinct}             \\
\label{rel9}
s_j\sigma_{ij} &=  \sigma_{i(j+1)} s_j                 &&
{\rm for \;} i,j,j+1 {\rm  \; distinct}                \\
\label{rel11} s_i \sigma_{ij}  &= \sigma_{(i+1)j}s_i && {\rm for
\;} i,i+1,j {\rm  \; distinct}
\end{align}
\end{alphalist}
\end{theorem}

\proof  Since the presentation ($a$) of $S_n$ is well-known, and Lin
\cite{Lin} proved that $PLB_n$ has the presentation ($b$), to
present their semidirect product $LB_n$ it suffices to add
relations that express the result of conjugating any of Lin's
generators $\sigma_{ij}$ by the symmetric group generators $s_k$.
For $1 \le i \le n-1$ we have:
\[
\def\objectstyle{\scriptstyle}
\def\labelstyle{\scriptstyle}
s_i \sigma_{i(i+1)}s_i^{-1}
\qquad =
\xy  0;<1cm,0cm>:
(-.4,2.6)*{i};
(.4,2.6)*{i+1};
(0,-1.5)*{\blor};
(0,0)*{\pltr};
(0,1.5)*{\brol};
\endxy
\quad
=
\quad
\xy  0;<1cm,0cm>:
(-.4,2.6)*{i};
(.4,2.6)*{i+1};
(0,-1.5)*{\dpipe};
(0,0)*{\prtl};
(0,1.5)*{\dpipe};
\endxy
\quad =
\sigma_{(i+1)i}
\]

\noindent
For $i,j,k$ and $k+1$ all distinct, we have:
\[
s_k  \sigma_{ij} s_k^{-1}
= \quad
\def\objectstyle{\scriptstyle}
\def\labelstyle{\scriptstyle}
\xy  0;<1cm,0cm>:
(-.9,2.6)*{i};
(-.3,2.6)*{j};
(.3,2.6)*{k};
(.9,2.6)*{k+1};
(-.6,-1.5)*{\dpipe};
(-.6,0)*{\pltr};
(-.6,1.5)*{\dpipe};
(.6,-1.5)*{\blor};
(.6,0)*{\dpipe};
(.6,1.5)*{\brol};
\endxy
\quad
=
\quad
\xy  0;<1cm, 0cm>:
(-.9,2.6)*{i};
(-.3,2.6)*{j};
(.3,2.6)*{k};
(.9,2.6)*{k+1};
(-.6,-1.5)*{\dpipe};
(-.6,0)*{\pltr};
(-.6,1.5)*{\dpipe};
(.6,-1.5)*{\dpipe};
(.6,0)*{\dpipe};
(.6,1.5)*{\dpipe};
\endxy
\quad =
\sigma_{ij}
\]
\noindent
For $i, j$ and $j+1$ distinct, we have:
\[
s_j \sigma_{ij} {s_j}^{-1} =
\def\objectstyle{\scriptstyle}
\def\labelstyle{\scriptstyle}
\qquad
\xy  0;<1cm,0cm>:
(-.6,2.6)*{i};
(0,2.6)*{j};
(.6,2.6)*{j+1};
(-.6,-1.5)*{\pipe};
(.3,-1.5)*{\blor};
(-.3,0)*{\pltr};
(.6,0)*{\pipe};
(-.6,1.5)*{\pipe};
(.3,1.5)*{\brol};
\endxy
\quad
= \sigma_{i(j+1)}
\]
and using a similar picture we see that for $i,i+1$ and $j$ distinct,
$s_i \sigma_{ij} {s_i}^{-1} = \sigma_{(i+1)j}$.  The reader may notice
that we have not included all possible conjugations of generators of $PLB_n$
by generators of $S_n$---we would naively expect two additional such classes,
yielding two more relations:
\begin{align}
\label{rel9prime}
  s_{j-1} \sigma_{ij}  &= \sigma_{i(j-1)}s_{j-1}                 &&
{\rm for \;} i,j-1,j {\rm  \; distinct} \\
\label{rel11prime}
  s_{i-1} \sigma_{ij}  &= \sigma_{(i-1)j}s_{i-1}                 &&
{\rm for \;} i-1,i,j {\rm  \; distinct}
\end{align}
but these follow, respectively, from (\ref{rel9}) and (\ref{rel11})
combined with (\ref{rel3}).
So, we have precisely the relations in part $(c)$, as desired.
\qed

  From this presentation of the loop braid group
we now derive a presentation with fewer generators.
We keep all the generators $s_i$, but replace the
$\sigma_{ij}$ with new generators
defined as follows:
\[        \sigma_i = s_i \sigma_{i(i+1)}  \]
for $1 \le i \le n-1$.  We can draw these as follows:
\[
\smalllabels
   \sigma_{i} =
\quad
   \xy 0;<1cm,0cm>:
      (0,-.75)*{\pltr};
      (-.35,1.75)*{i};
      (.35,1.75)*{i+1};
      (0,.75)*{\brol};
    \endxy
\quad
=
\quad
   \xy 0;<1cm,0cm>:
      (0,0)*{\brtl};
      (-.35,1)*{i};
      (.35,1)*{i+1};
    \endxy
\]
where we twist the picture a bit in the second step.  To see
that the generators $s_i$ and $\sigma_i$ indeed give a new
presentation, note that we can express the old generators
$\sigma_{ij}$ in terms of these new ones as follows.  First, repeatedly
applying (\ref{rel9}) we obtain:
\[
\begin{array}{cl}
 \sigma_{ij}=   s_{j-1}s_{j-2} \cdots   s_{i+1}
                        \sigma_{i(i+1)}
                        s_{i+1} s_{i+2}\cdots s_{j-2} s_{j-1}
& \textrm{for \;} i < j.
\end{array}
\]
If instead of (\ref{rel9}) we use its equivalent form
(\ref{rel9prime}), we obtain:
\[
\begin{array}{cl}
 \sigma_{ij}=      s_j s_{j+1} \cdots s_{i-2}
                         \sigma_{i(i-1)}
                         s_{i-2} \cdots s_{j+1} s_j
& \textrm{for \;} i > j.
\end{array}
\]
Rewriting these in terms of the new generators $\sigma_i$,
and in the second case using relation (\ref{rel7}), we obtain a
way to write $\sigma_{ij}$ in terms of the new generators:
\begin{equation}
\label{generatorCorrespondence1}
   \sigma_{ij}
=
\left\{
\begin{array}{cl}
    s_{j-1}s_{j-2} \cdots   s_{i}
                        \sigma_{i}
                        s_{i+1} s_{i+2}\cdots s_{j-2} s_{j-1}
& \textrm{for \;} i < j \\
   s_j s_{j+1} \cdots s_{i-2}
                         \sigma_{i-1}
                         s_{i-1} s_{i-2} \cdots s_{j+1} s_j
& \textrm{for \;} i > j
\end{array}
\right.
\end{equation}
Sometimes it is more convenient to use an alternate formula, obtained
by applying (\ref{rel11}), its equivalent form (\ref{rel11prime}),
and (\ref{rel7}) again:
\begin{equation}
\label{generatorCorrespondence2}   \sigma_{ij}
=
\left\{
\begin{array}{cl}
s_i s_{i+1} \cdots s_{j-1} \sigma_{j-1} s_{j-2} \cdots
s_{i+1}s_i & \textrm{for \;} i < j \\
s_{i-1}s_{i-2} \cdots s_{j+1} \sigma_{j}s_js_{j+1} \cdots  s_{i-2} s_{i-1} & \textrm{for \;} i > j.
\end{array}
\right.
\end{equation}
What these formulas say is that when $j\neq i+1$ we can
construct the loop braid $\sigma_{ij}$ by permuting either
the $i$th circle or the
$j$th until they are adjacent, braiding one through the other,
and then permuting the circles back to where they started.

The nice thing about using $s_i$ and $\sigma_i$ as generators
of the loop braid group is that
$s_i$ describes how two neighboring circles
can trade places by going around each other:
\[
\smalllabels
 s_i =
   \xy 0;<1cm,0cm>:
      (0,0)*{\brol};
      (-.35,1)*{i};
      (.35,1)*{i+1};
    \endxy
\]
\noindent
while $\sigma_i$ describes how two neighboring circles
can trade places with the right one passing over and then down
through the left one:
\[
\smalllabels
\sigma_i =
   \xy 0;<1cm,0cm>:
      (0,0)*{\brtl};
      (-.35,1)*{i};
      (.35,1)*{i+1};
    \endxy
\]
As a result, the generators $s_i$ generate a subgroup of
$LB_n$ isomorphic to the symmetric group $S_n$, while
the $\sigma_i$ generate a subgroup isomorphic to the braid
group $B_n$.  There are also `mixed relations' involving generators
of both kinds:

\begin{theorem} \et \label{presentation2}
The loop braid group $LB_n$ has a
presentation with generators $s_i$ and $\sigma_i$ for $1 \le i \le n-1$
together with the following relations:

\begin{itemize}

\item[(a)]
relations for the standard generators $s_i$ of $S_n$:
\begin{align}
\label{Rel1}
s_i s_j &= s_j s_i                      &&
{\rm for \; } |i-j| > 1                 \\
\label{Rel2}
s_i s_{i+1} s_i &= s_{i+1} s_i s_{i+1}  &&
{\rm for \; } 1 \le i \le n-2           \\
\label{Rel3}
s_i^2 &= 1                              &&
{\rm for \; } 1 \le i \le n-1
\end{align}

\item[($b^{\prime}$)]
relations for the standard generators $\sigma_i$ of $B_n$:
\begin{align}
\label{Rel4}
\sigma_i \sigma_j &= \sigma_j \sigma_i                                &&
{\rm for \; } |i-j| > 1                                               \\
\label{Rel5}
\sigma_i \sigma_{i+1} \sigma_i &= \sigma_{i+1} \sigma_i \sigma_{i+1}  &&
{\rm for \; } 1 \le i \le n-2
\end{align}

\item[($c^{\prime}$)] the following {\bf mixed relations}:
\begin{align}
\label{Rel7} s_i \sigma_j &= \sigma_j s_i &&
{\rm for \; } |i-j| > 1                                               \\
\label{Rel8} s_i s_{i+1} \sigma_i &= \sigma_{i+1} s_i s_{i+1} &&
{\rm for \; } 1 \le i \le n-2                                         \\
\label{Rel9} \sigma_i \sigma_{i+1} s_i &= s_{i+1} \sigma_i
\sigma_{i+1}            && {\rm for \; } 1 \le i \le n-2
\end{align}

\end{itemize}
\end{theorem}

\proof  The proof is somewhat lengthy, so we defer
it to the Appendix.  It is, however, simple to convince oneself using
pictures that the given relations express topologically allowed moves
for loop braids.  Perhaps the least obvious
of these is (\ref{Rel9}), for which we supply a visual proof below:
\[
\xy  0;<1cm,0cm>: (-.3,-1.5)*{\brol}; (.6,-1.5)*{\pipe};
(-.6,0)*{\pipe}; (.3,0)*{\brtl}; (-.3,1.5)*{\brtl};
(.6,1.5)*{\pipe};
\endxy
\quad = \quad \xy
 (0,0)*{\includegraphics{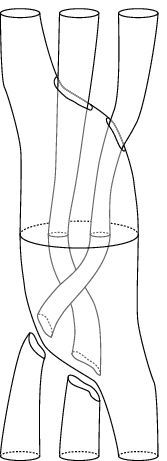}}
\endxy
\quad = \quad \xy
 (0,0)*{\includegraphics{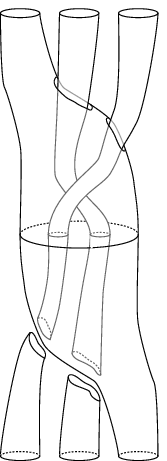}}
\endxy
\quad = \quad \xy  0;<1cm,0cm>: (-.6,-1.5)*{\pipe};
(.3,-1.5)*{\brtl}; (-.3,0)*{\brtl}; (.6,0)*{\pipe};
(-.6,1.5)*{\pipe}; (.3,1.5)*{\brol};
\endxy
\]
\qed

\newpage

If we omit relations (\ref{Rel9}) we obtain the
`virtual braid group' $VB_n$ of Vershinin \cite{Vershinin}.
This plays a role in virtual knot theory analogous to that
of the usual braid group in ordinary knot theory.
If we include these relations, which say:
\[
\xy  0;<1cm,0cm>:
(-.3,-1.5)*{\brol};
(.6,-1.5)*{\pipe};
(-.6,0)*{\pipe};
(.3,0)*{\brtl};
(-.3,1.5)*{\brtl};
(.6,1.5)*{\pipe};
\endxy
\quad
=
\quad
\xy  0;<1cm,0cm>:
(-.6,-1.5)*{\pipe};
(.3,-1.5)*{\brtl};
(-.3,0)*{\brtl};
(.6,0)*{\pipe};
(-.6,1.5)*{\pipe};
(.3,1.5)*{\brol};
\endxy
\]
then we obtain precisely the
`braid permutation group' $BP_n$ of Fenn, Rim\'anyi and Rourke \cite{FRR}.
So, the loop braid group is isomorphic to the braid permutation group.

The isomorphism $LB_n \iso BP_n$ yields a simplified diagrammatic way of
working with loop braids, which is in fact the method used by
Fenn, Rim\'anyi and Rourke in their original paper on $BP_n$.  In the theory of
`welded braids', the generators $\sigma_i$ in $BP_n$ correspond to the kind of
crossings found in ordinary braids:
$\xy \knotholesize{4pt}\xoverv~{(-1,1.3)}{(1,1.3)}{(-1,-1.3)}{(1,-1.3)};\endxy$,
while the $s_i$ describe `welded crossings', drawn like this:
$\xy
{\knotholesize{0pt}}
\xoverv~{(-1,1.3)}{(1,1.3)}{(-1,-1.3)}{(1,-1.3)};(0,0)*{\scriptstyle \bullet}
\endxy$.
These crossings are called `welded' because one imagines that the two strands
have been `welded down' at the crossing.  The point is that elements of the
abstract group presented in Theorem \ref{presentation2} can be represented
either as loop braid diagrams or as welded braid diagrams, as follows:

\[
\xy
(-20,0)*{\xy
    (-19,0)*{s_i=};
    (7,0)*{\smalllabels\xy0;<1cm,0cm>:
            \zbendv~{(0,.75)}{(.4,.75)}{(0,-.75)}{(.4,-.75)};
            \sbendv~{(0,.75)}{(.4,.75)}{(0,-.75)}{(.4,-.75)};
            (.2,0)*{\textstyle \bullet};
            (0,1)*{i};(0,-1)*{};
            (.5,1)*{i+1};
          \endxy};
    (0,0)*{=};
    (-10,0)*{\smalllabels \xy 0;<1cm,0cm>:
                (0,0)*{\brol};
                (-.35,1)*{i};
                (.35,1)*{i+1};
                (-.35,-1)*{};
                (.35,-1)*{};
            \endxy};
     \endxy};
(20,0)*{\xy
    (-19,0)*{\sigma_i=};
    (7,0)*{\smalllabels\xy0;<1cm,0cm>:
            \vcross~{(0,.75)}{(.4,.75)}{(0,-.75)}{(.4,-.75)};
            (0,1)*{i};(0,-1)*{};
            (.5,1)*{i+1};
          \endxy};
    (0,0)*{=};
    (-10,0)*{\smalllabels \xy 0;<1cm,0cm>:
                (0,0)*{\brtl};
                (-.35,1)*{i};
                (.35,1)*{i+1};
                (-.35,-1)*{};
                (.35,-1)*{};
            \endxy};
     \endxy};
\endxy
\]
For the {\em pure} loop braid group $PLB_n$, the above correspondence 
implies the following welded braid pictures of the generators
$\sigma_{i(i+1)}$ and their inverses:
\[
\xy
(-25,0)*{\xy
    (-20,0)*{{\sigma_{i(i+1)}}^{\phantom{\;1}}};     (-15,0)*{=};
    (-7.5,0)*{\smalllabels \xy 0;<1cm,0cm>:
                (0,0)*{\pltr};
                (-.35,1)*{i};
                (.35,1)*{i+1};
                (-.35,-1)*{};
                (.35,-1)*{};
            \endxy};
    (0,0)*{=};
    (5,0)*{\xy0;<1cm,0cm>:
            \zbendv~{(0,.75)}{(.4,.75)}{(0,0)}{(.4,0)};
            \sbendv~{(0,.75)}{(.4,.75)}{(0,0)}{(.4,0)};
            (.2,.375)*{\bullet};
            \vcross~{(0,0)}{(.4,0)}{(0,-.75)}{(.4,-.75)};
          \endxy};
   \endxy};
(25,0)*{\xy
    (-22,0)*{{\sigma_{i(i+1)}}^{-1}=};
    (-7.5,0)*{\smalllabels \xy 0;<1cm,0cm>:
                (0,0)*{\pltrinv};
                (-.35,1)*{i};
                (.35,1)*{i+1};
                (-.35,-1)*{};
                (.35,-1)*{};
            \endxy};
    (0,0)*{=};
    (5,0)*{\xy0;<1cm,0cm>:
            \vtwist~{(0,.75)}{(.4,.75)}{(0,0)}{(.4,0)};
            (.2,-.375)*{\bullet};
            \zbendv~{(0,0)}{(.4,0)}{(0,-.75)}{(.4,-.75)};
            \sbendv~{(0,0)}{(.4,0)}{(0,-.75)}{(.4,-.75)};
          \endxy};
   \endxy};
\endxy
\]
The other generators $\sigma_{ij}$ can be obtained from these by conjugation,
using (\ref{generatorCorrespondence1}) or (\ref{generatorCorrespondence2}).
For example:
\[
\xy
(25,0)*{\xy
    (-20,0)*{{\sigma_{(i+1)i}}=};
    (-7.5,0)*{\smalllabels \xy 0;<1cm,0cm>:
                (0,0)*{\prtl};
                (-.35,1)*{i};
                (.35,1)*{i+1};
                (-.35,-1)*{};
                (.35,-1)*{};
            \endxy};
    (0,0)*{=};
    (5,0)*{\xy0;<1cm,0cm>:
            \vcross~{(0,.75)}{(.4,.75)}{(0,0)}{(.4,0)};
            (.2,-.375)*{\bullet};
            \zbendv~{(0,0)}{(.4,0)}{(0,-.75)}{(.4,-.75)};
            \sbendv~{(0,0)}{(.4,0)}{(0,-.75)}{(.4,-.75)};
          \endxy};
   \endxy};
\endxy
\]

Diagrammatic calcuations with welded braids---and hence with loop braids---can
be carried out by using the usual Reidemeister moves for real crossings,
along with `welded Reidemeister moves':
\[  
\xy
(20,10)*{\includegraphics{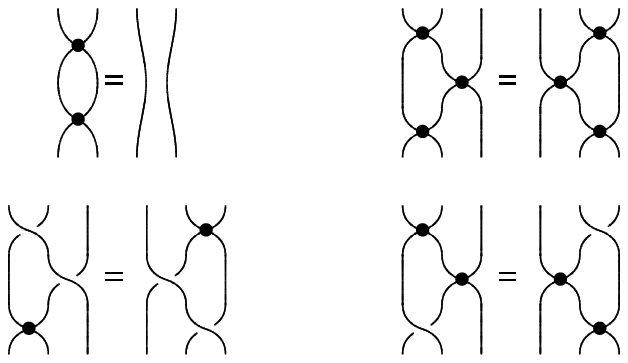}};
\endxy
\]
which are of course simply graphical restatements of the relations in
$(a)$ and $(c^\prime)$.  The nonexistence of the following move:
\[  
\xy
(-7,0)*{\xy 0;<1cm,0cm>:
           {\knotholesize{0pt}}
           \vcross~{(0,.75)}{(.4,.75)}{(0,.25)}{(.4,.25)};
            (.2,.5)*{\bullet};
            \sbendv~{(.8,.75)}{(.8,.25)}{(.8,.75)}{(.8,.25)};
            {\knotholesize{5pt}}
           \vcross~{(.4,.25)}{(.8,.25)}{(.4,-.25)}{(.8,-.25)};
            \sbendv~{(0,.25)}{(0,-.25)}{(0,.25)}{(0,-.25)};
            \vcross~{(0,-.25)}{(.4,-.25)}{(0,-.75)}{(.4,-.75)};
            \sbendv~{(.8,-.25)}{(.8,-.25)}{(.8,-.75)}{(.8,-.75)};
        \endxy};
(0,0)*{\neq};
(7,0)*{\xy 0;<1cm,0cm>:
            \vcross~{(0,.75)}{(.4,.75)}{(0,.25)}{(.4,.25)};
            \sbendv~{(-.4,.75)}{(-.4,.25)}{(-.4,.75)}{(-.4,.25)};
            \vcross~{(-.4,.25)}{(0,.25)}{(-.4,-.25)}{(0,-.25)};
            {\knotholesize{0pt}}
            \sbendv~{(.4,.25)}{(.4,-.25)}{(.4,.25)}{(.4,-.25)};
            \vcross~{(0,-.25)}{(.4,-.25)}{(0,-.75)}{(.4,-.75)};
            (.2,-.5)*{\bullet};
            \sbendv~{(-.4,-.25)}{(-.4,-.25)}{(-.4,-.75)}{(-.4,-.75)};
        \endxy};
\endxy
\]
is the rationale for the term `welded braid'---we are not allowed
to pass a strand under the weld.

It is easy from the presentation in Theorem \ref{presentation2} to work out
the 1-dimensional unitary representations of the loop braid group.
If $\rho \maps LB_n \to \U(1)$ is such a representation, we must have
\[   \rho(s_i) = \pm 1  \]
and
\[   \rho(\sigma_i) = q  \]
for all $1 \le i < n$, where $q \in \U(1)$ is a fixed phase.
We call the representations with $\rho(s_i) = 1$ {\bf bose-anyons},
and the representations with $\rho(s_i) = -1$ {\bf fermi-anyons}.
These have been studied in physics at least since the work of 
Balachandran \cite{Balachandran}, and recently Niemi has shown how they
arise in the dynamics of vortices in a quantum fluid \cite{Niemi}.

In Section \ref{4d} we describe more interesting unitary
representations of the loop braid group, using some technology
which we now develop.  In related work, Szabo \cite{Szabo} has obtained
a different class of representations using $BF$ theory with abelian 
gauge group.  Surya \cite{Surya} has also studied representations of 
the loop braid group.

\section{Motion Groups and Flat Bundles}
\label{motion}

In this section we recall Dahm's \cite{Dahm} action of the
motion group $\Mo(S,\Sigma)$ on the fundamental group of $S - \Sigma$
and describe how this gives a unitary representation of the motion
group on a certain Hilbert space of states for $BF$ theory on
$S - \Sigma$.

We consider $BF$ theory in $n$-dimensional spacetime.  So, we
take `space' to be of the form $X = S - \Sigma$,
where $S$ is an oriented manifold of dimension $n-1$, and $\Sigma
\subset S$ is an oriented submanifold.
We let $G$ be a Lie group and let $P \to X$ be a principal $G$-bundle.
The `naive configuration space' of $BF$ theory
is $\A_0/\G$, where $\A_0$ is the space of flat connections
on $P$ and $\G$ is the group of gauge transformations.
By `naive' we mean that we are ignoring boundary conditions;
there are no boundary conditions to worry about when $X$ is compact, but
we shall mainly be interested in two examples where it is not:

\begin{enumerate}
\item
$X$ is $\R^2$ with a finite set of points
removed (describing {\it point particles}):
\[   X = S - \Sigma, \qquad S = \R^2, \qquad
\Sigma = \{z_1, \dots, z_n\}. \]
\item
$X$ is $\R^3$ with a finite set of
unlinked unknotted circles removed (describing what one might call
{\it closed strings}):
\[   X = S - \Sigma, \qquad S = \R^3, \qquad
     \Sigma = \ell_1 \cup \cdots \cup \ell_n. \]
\end{enumerate}

\noindent A rigorous study of $BF$ theory may require that
we impose boundary conditions at $\Sigma$.  We ignore this issue
now, leaving it for future research.

The space $\A_0/\G$ is a bit difficult to handle.
It is often more convenient to start by fixing a basepoint
$\ast \in X$ and working with $\A_0/\G_0$, where
\[         \G_0 = \{g \in \G \colon \; g(\ast) = 1 \}  .\]
The group $\G/\G_0 \iso G$ acts on $\A_0/\G_0$ in a natural
way.  This lets us form $\A_0/\G$ as the quotient of the
bigger space $\A_0/\G_0$ by this action of $G$.

The advantage of the space $\A_0/\G_0$ is that any point
$[A]$ in this space gives a homomorphism
\[  \hol([A]) \maps \pi_1(X) \to G  \]
which sends any homotopy class of loops $[\gamma]$ to the
holonomy of $A$ around $\gamma$.  This gives a map
\[    \hol \maps \A_0/\G_0 \to \hom(\pi_1(X), G)   \]
which is known to be one-to-one.
Note that $G$ acts on $\hom(\pi_1(X),G)$ by conjugation:
\[             (gf)(\gamma) = gf(\gamma)g^{-1}  \]
where $f \maps \pi_1(X) \to G$ is any homomorphism.
Moreover, the map $\hol$ is compatible with this group action:
\[    \hol([gA]) = g \, \hol([A])   .\]

So far we have fixed a principal $G$-bundle $P$.  But, in gauge theory it
is often better to treat this bundle as variable---part of the physical
field along with the connection $A$.  For example, path integrals
in quantum chromodynamics involve a sum over bundles, which represent
instantons.   The mathematical advantage of treating $P$ as variable is
that all points of $\hom(\pi_1(X),G)$ are in the image of $\hol$
if we allow ourselves to vary $P$ \cite{Kobayashi}.
A point in this space represents a `$G$-bundle
with flat connection over $X$, mod gauge transformations
that equal the identity at the basepoint'.  Modding out by the
rest of the gauge transformations we get a space known
as the `moduli space of flat bundles', $\hom(\pi_1(X),G)/G$.
This is the naive configuration space for $BF$
theory where we treat the bundle $P$ as variable.

Applying Schr\"odinger quantization to this configuration
space, we obtain the (naive) Hilbert space for $BF$ theory:
\[           L^2(\hom(\pi_1(X),G)/G)   \]
Of course, defining this $L^2$ space
requires that we choose a measure on the moduli space
of flat bundles.  Alternatively, we can try to form a Hilbert space
\[           L^2(\hom(\pi_1(X),G))    \]
on which $G$ acts as follows:
\[            (g\psi)(f) = \psi (g^{-1}f)  .\]
Again, this requires choosing a measure on $\hom(\pi_1(X),G)$.
Moreover, $G$ will only have a unitary representation on
$L^2(\hom(\pi_1(X),G)$ if this measure is $G$-invariant.

In Sections \ref{3d} and \ref{4d}
we will show that for the two examples above,
there is a `natural' choice of $G$-invariant measure on
$\hom(\pi_1(X),G)$.
In both these examples the motion group $\Mo(S,\Sigma)$
acts on $\pi_1(X)$ and thus on $\hom(\pi_1(X),G)$.
By saying a measure on $\hom(\pi_1(X),G)$
is `natural', we simply mean that it is preserved by this action.

Using such a natural measure to define the Hilbert space
$L^2(\hom(\pi_1(X),G))$, we obtain a unitary representation
of the motion group on this Hilbert space.  This representation
describes the statistics of point particles or closed strings in $BF$ theory.
As we have seen, in the first example the motion
group is the braid group $B_n$, while in the 4d case it is the
loop braid group $LB_n$.  So, we obtain `exotic statistics' in both
cases.  This fact is somewhat familiar in 3 dimensions, but less
so in 4 dimensions.  So, in the following sections we
first review the 3d case, and then move on to the 4d case
after a brief digression on `quandle field theory'.

Before doing this, however, let us see how the motion group acts on
$\pi_1(X)$.  The idea goes back to Dahm's original work on
the motion group \cite{Dahm}, and it has been nicely explained 
by Goldsmith \cite{Goldsmith}.  The idea is simple:
elements of the motion group $\Mo(S,\Sigma)$ give
equivalence classes of diffeomorphisms of $X = S - \Sigma$, and these
act on homotopy classes of loops in $X$.   The only
problem is that the fundamental group is defined using {\it based}
loops, and the diffeomorphisms used in the definition of the
motion group need not preserve the basepoint in $X$.
Luckily, Wattenberg \cite{Wattenberg} has shown that we can use
compactly supported diffeomorphisms in the definition of the motion group
without changing this group.  In the examples above, we can assume without
loss of generality that these diffeomorphisms are supported in a fixed large
ball containing $\Sigma$.  So, if we choose a
basepoint $\ast \in S$ that is sufficiently far from $\Sigma$,
we can assume this basepoint is preserved by all the diffeomorphisms
in the definition of the motion group.  This makes it easy to check
that $\Mo(S,\Sigma)$ acts as automorphisms of $\pi_1(X)$.

{\boldmath
\section{Point Particles in 3d $BF$ Theory}
}
\label{3d}

Now let us apply the general ideas of the previous section to
the case of a plane with $n$ punctures:
\[
X = S - \Sigma, \qquad S = \R^2, \qquad \Sigma = \{z_1, \dots, z_n\}
\]
If we interpret these punctures as `particles',
we shall see that a state of 3d $BF$ theory on this space
describes a collection of identical point particles
with exotic statistics governed by the braid group.

The fundamental group of $X$ is the free group on $n$ generators,
so we have
\[
\hom(\pi_1(X),G) = G^n
\]
The $n$ group elements here are nothing but the holonomies of a
flat connection around based loops going clockwise
around the particles:
\[
\smalllabels
\xy 0;<1cm,0cm>:
(0,0)*{\includegraphics{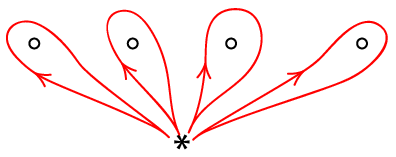}};
(1,.3)*{\cdots};
(-1.7,.7)*{g_1};
(-.7,.8)*{g_2};
(.3,.8)*{g_3};
(1.6,.7)*{g_n};
\endxy
\]
Having described particles as punctures in this theory, let us now
consider what sort of statistics such particles obey.
The previous section shows that the interchange of identical
particles is described by an action of the $n$-strand braid
group $B_n$ on $G^n$, but we would like to work it out explicitly.
For simplicity, consider the case $n = 2$ and consider what
happens when the two particles switch places.  As remarked earlier,
there are infinitely many topologically distinct ways for the particles
to move around each other, but they are all powers of
the braid group generator $\sigma_1$:
\[
\def\objectstyle{\scriptstyle}
\def\labelstyle{\scriptstyle}
\xy
(-20,0)*{\xy
   \vcross~{(-5,10)}{(5,10)}{(-5,-10)}{(5,-10)};
    (-5,12)*{};
    (5,12)*{};
    (-5,-12)*{};
    (5,-12)*{};
\endxy};
\endxy
\]
If the holonomies around the two particles are $g_1$
and $g_2$:
\[
\smalllabels
\xy
(-2,-1.5)*{\includegraphics{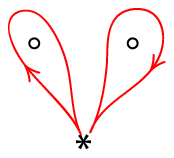}};
(-6,7)*{g_1};
(6,7)*{g_2};
\endxy
\]
switching them via $\sigma_1$ induces a diffeomorphism
of the plane which deforms the loops around which the holonomies are taken:
\[
\xy
(0,0)*{\includegraphics{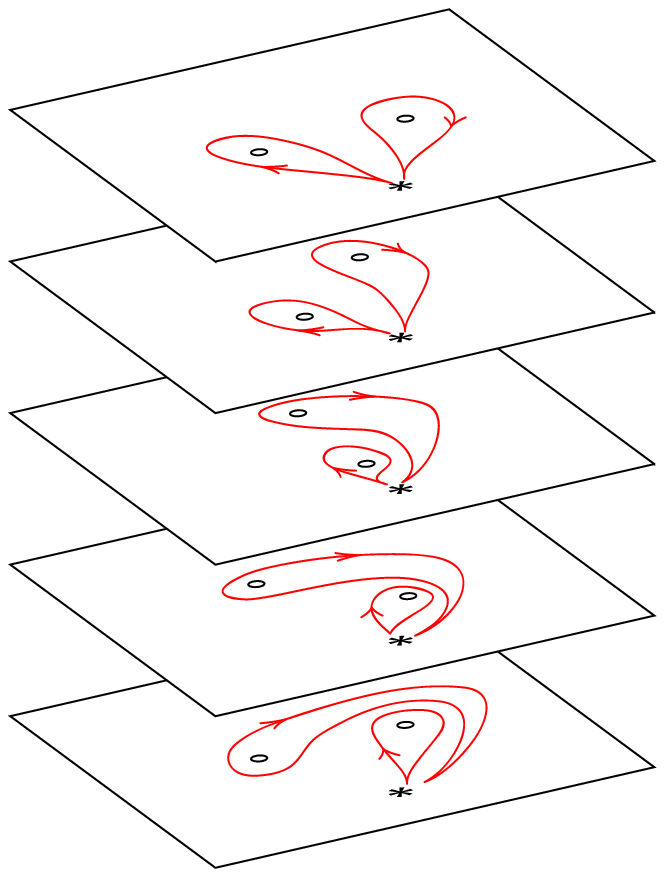}};
\endxy
\]
To see how the system changes in this process, compare the final
frame in this `movie' to the first frame.  Given that $(g_1,g_2)\in G^2$
describes the holonomies initially, a slight deformation of the loops
in the final frame:
\[
\smalllabels
\xy
(0,-.5)*{\includegraphics{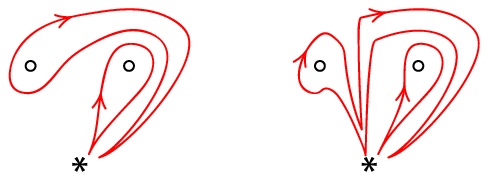}};
(0,0)*{\textstyle =};
(-21,7.5)*{g_2};
(-16.5,0)*{g_1};
(9,7)*{g_2};
(15,1)*{g_1};
\endxy
\]
makes it clear that the corresponding holonomies
around {\em these} loops in the final configuration:
\[
\smalllabels
\xy
(-2,-1.5)*{\includegraphics{bgact2a.ps}};
(-6,7.7)*{g'_1};
(6,7.7)*{g'_2};
\endxy
\]
is $(g'_1,g'_2)=(g_1 g_2 {g_1}^{\!-1},g_1)$.  Thus the effect of switching
the two particles via $\sigma_1$ is to send $(g_1,g_2)$ to
$(g_1 g_2 {g_1}^{\!-1},g_1)$.

We can work out the action of $\sigma_1^{-1}$ in the same way,
or simply derive it algebraically from the fact that it must undo
the effect of $\sigma_1$.
The easiest way to remember the results is with this picture:
\[
\def\objectstyle{\scriptstyle}
\def\labelstyle{\scriptstyle}
\xy
(-20,0)*{\xy
 \vcross~{(-5,10)}{(5,10)}{(-5,-10)}{(5,-10)};
    (-5,12)*{g_1};
    (5,12)*{g_2};
    (-5,-12)*{g_1 g_2 g_1^{-1}};
    (5,-12)*{g_1};
\endxy};
(20,0)*{\xy
   \vtwist~{(-5,10)}{(5,10)}{(-5,-10)}{(5,-10)};
    (-5,12)*{g_1};
    (5,12)*{g_2};
    (-5,-12)*{g_2};
    (5,-12)*{g_2^{-1} g_1 g_2};
\endxy};
\endxy
\]

More generally, we have a right action of the braid group $B_n$ on
$G^n$ given as follows:
\[   (g_1, \dots, g_i, g_{i+1}, \dots, g_n)\sigma_i  =
(g_1, \dots, g_{i} g_{i+1} g_{i}^{-1}, g_i, \dots, g_n)  .\]
As mentioned in the previous section, we also have a left action of
$G$ on $G^n$ via gauge transformations at the basepoint $\ast$.
This works as follows:
\[    g(g_1, \dots, g_n) = (gg_1g^{-1}, \dots, gg_ng^{-1})  .\]

We would like a measure on $G^n$ that is invariant under both
these group actions, so that the braid group and gauge transformations
act as unitary operators on $L^2(G^n)$.  Such a measure exists
whenever $G$ is {\bf unimodular}, meaning that its
left-invariant Haar measure is also right-invariant.
A Lie group is automatically unimodular if it is compact, or
abelian, or semisimple.  In particular, the groups $\SO(p,q)$
are all unimodular.   Since these groups act on Minkowski spacetime
in a way that preserves its Lebesgue measure, the Poincar\'e groups
$\ISO(p,q)$ are also unimodular.  Also, the identity component of
a unimodular group is unimodular, as is any covering space of a
unimodular group.

  From this we see that the 3d Lorentz group $\SO(2,1)$ is unimodular,
as are its identity component $\SO_0(2,1)$ and the double cover of its
identity component, namely $\SL(2,\R)$.  All these are reasonable
choices of gauge group when treating 3-dimensional---or more properly,
(2+1)-dimensional---Lorentzian gravity as a $BF$ theory.

Given a unimodular Lie group, Haar measure is typically not the
only measure invariant under conjugation: we can multiply Haar
measure by any function that only depends on the conjugacy class.
As an extreme example, we can even try to multiply Haar measure by
a `delta function' supported on one conjugacy class.  More
precisely, we can look for a conjugation-invariant measure
supported on a single conjugacy class of $G$.  In this case we
might as well be working not with $G$ but with just the conjugacy
class.  It turns out that in the case of 3d quantum gravity, this
amounts to studying identical particles {\it of a specified mass}.
This leads us to our next subject: quandle field theory.

\hskip 1em

\section{Quandle Field Theory}
\label{quandle}

In the previous section we considered $BF$ theory in 3 dimensions,
and were led to a natural action of the braid group $B_n$ on the
space $G^n$ for any group $G$.  Notice that we did not actually need
the multiplication in $G$ to define this action; we only needed
the operation of {\it conjugation}.  This suggests that we can
work more generally, replacing the group $G$ by some algebraic
structure that captures the properties of conjugation.  Such a
thing is called a `quandle'.

More precisely, a {\bf quandle} is a nonempty set $Q$ equipped with
two binary operations $\zap \maps Q \times Q \to Q$ and
$\paz \maps Q \times Q \to Q$ called {\bf left} and {\bf right conjugation},
which satisfy:
\begin{description}
\item[(i)] left idempotence: $x \rhd x = x$
\item[(i$^\prime$)] right idempotence: $x \lhd x = x$
\item[(ii)] left inverse law: $x \rhd (y \lhd x) = y$
\item[(ii$^\prime$)] right inverse law: $(x \rhd y) \lhd x = y$
\item [(iii)] left distributive law:
$x \rhd (y \rhd z) = (x \rhd y) \rhd (x \rhd z)$
\item[(iii$^\prime$)] right distributive law:
$(x \lhd y ) \lhd z = (x \lhd z) \lhd (y \lhd z)$
\end{description}
for all $x, y, z \in Q$.
In general, the operations of left and right conjugation in a quandle
are neither associative nor commutative.

Quandles were first introduced as a source of knot invariants by David
Joyce \cite{J} in $1982$.
Many examples of quandles can be found in the work of Fenn
and Rourke \cite{FR} and other authors \cite{Brieskorn, J, K}.
For us, the most important examples come
from taking a group $G$, letting $Q$ be any union of conjugacy
classes of $G$, and making $Q$ into a quandle with
\[             g \zap h = ghg^{-1}   , \qquad h \paz g = g^{-1}hg . \]
We are especially interested in the case where $Q$ is either the whole
group $G$ or a single conjugacy class.

We can do some of the same things with quandles as with groups.
For example, we can define a {\bf topological quandle} to be a
topological space that is also a quandle in such a way that
the quandle operations $\zap$ and $\paz$ are continuous
\cite{Rubinsztein2}.  If $G$ is a Lie group and $Q \subseteq G$
is a conjugacy class, $Q$ becomes a topological quandle with
the induced topology.

Given a topological quandle $Q$, we define an {\bf invariant measure}
on $Q$ to be a Borel measure that is invariant under left conjugation
by any element of $Q$---or equivalently, invariant under right conjugation
by any element of $Q$.  This implies that
\[
\begin{array}{ccl}
   \displaystyle{\int} f(x)\; d\mu(x) &=&
   \displaystyle{\int} f(q\zap x)\; d\mu(x) \\  \\
   &=& \displaystyle{\int} f(x \paz q)\; d\mu(x)
\end{array}
\]
for any $q\in Q$ and any integrable function $f$ on $Q$.
As noted earlier, invariant measures on quandles are far
from unique in general.  In particular, we may multiply an
invariant measure on a Lie group by any class function and obtain a
new invariant measure.

In the previous section, we saw that the $n$-strand braid group $B_n$
acts on $G^n$ for any group $G$.  But, since our argument relied only on
properties of conjugation, it works just as well for a quandle.  The idea
is that we can braid two elements of a quandle past each other using
left conjugation:
\[
\def\objectstyle{\scriptstyle}
\def\labelstyle{\scriptstyle}
{\xy
 \vcross~{(-5,10)}{(5,10)}{(-5,-10)}{(5,-10)};
    (-5,12)*{x};
    (5,12)*{y};
    (-5,-12)*{x\zap y};
    (5,-12)*{x};
\endxy}
\]
The inverse braiding uses right conjugation:
\[
\def\objectstyle{\scriptstyle}
\def\labelstyle{\scriptstyle}
{\xy
   \vtwist~{(-5,10)}{(5,10)}{(-5,-10)}{(5,-10)};
    (-5,12)*{x};
    (5,12)*{y};
    (-5,-12)*{y};
    (5,-12)*{x\paz y};
\endxy}
\]
It is well known that with these rules, the
braid group relations follow from the quandle axioms.
So, generalizing our result from the previous section, we easily
obtain:

\begin{theorem} \label{3dtheorem} \et
Suppose $Q$ is a topological quandle equipped
with an invariant measure.  Then there is a unitary representation
$\rho$ of the braid group $B_n$ on $L^2(Q^n)$ given by
\[         (\rho(\sigma) \psi)(q_1,\dots, q_n) =
                \psi((q_1, \dots, q_n)\sigma) \]
for all $\sigma \in B_n$, where $B_n$ has a right action on $Q^n$
given by:
\[         (q_1, \dots, q_i, q_{i+1}, \dots, q_n)\sigma_i =
(q_1, \dots, q_i \zap q_{i+1}, q_i, \dots, q_n)  .\]
There is also a unitary operator $U(q)$ on $L^2(Q^n)$ for
each element $q \in Q$, given by
\[     (U(q)\psi)(q_1, \dots, q_n)  =
    \psi(q \zap q_1, \dots, q \zap q_n)  .\]
\end{theorem}

\noindent
The operators $U(q)$ represent gauge transformations when $Q$ is
a group, so we can think of them as representing some sort of
`gauge transformation' even when $Q$ is a quandle.  Of course, if
$Q$ is a conjugacy class in a group $G$, there will be
gauge transformations even for elements of $G$ that do not lie in $Q$.

It is instructive to work out the details in the case of
(2+1)-dimensional quantum gravity.  This theory can be viewed as a
$BF$ theory with $G$ being the connected Lorentz group
$\SO_0(2,1)$, or perhaps better, its double cover $\SL(2,\R)$.  In
either case we shall see that different conjugacy classes $Q$
describe different types of spinless particles.  The Hilbert space
for $n$ particles of this type is $L^2(Q^n)$, and Theorem
\ref{3dtheorem} describes the exotic statistics and gauge
invariance of this $n$-particle system.

In quantum field theory without gravity on 3d Minkowski spacetime, we
can describe the energy-momentum of a particle by an element $p \in \Sl(2,\R)$:
\[
   p = \left(
     \begin{array}{cc}
        p_x &  p_y+E \\
        p_y-E & -p_x
     \end{array}
    \right)
\]
Note that
\[
      \det p = E^2-p_x^2-p_y^2.
\]
The adjoint action of $\SL(2,\R)$ on its Lie algebra:
\begin{align*}
      \SL(2,\R) \times \Sl(2,\R) &\to \Sl(2,\R) \\
            (g,p) &\mapsto  gpg^{-1}
\end{align*}
preserves the determinant of $p$.
So, the adjoint action gives an action of $\SL(2,\R)$
as Lorentz transformations on the space of energy-momenta.
As explained in the Introduction, an orbit of this action
is just a type of spin-zero particle.

When we turn on gravity, we must describe energy-momenta not
by elements of the Lie algebra $\Sl(2,\R)$ but by elements of
the group $\SL(2,\R)$.  Particle types are then described not
by adjoint orbits but by conjugacy classes $Q \subseteq \SL(2,\R)$.
However, this new description is compatible with the old one, at least
for energy-momenta that are small compared to the Planck energy
$2\pi/\kappa$.  The reason is that we can identify group
elements near the identity with Lie algebra elements via the map
\[
\begin{array}{ccl}
          \Sl(2,\R) &\to& \SL(2,\R)   \\
              p    &\mapsto& \exp(\kappa p)
\end{array}
\]
This maps any adjoint orbit of $\Sl(2,\R)$ into a conjugacy class
of $\SL(2,\R)$.  Indeed, it gives a one-to-one
correspondence between the set of adjoint orbits close to $0 \in \Sl(2,\R)$
and the set of conjugacy classes close to $1 \in \SL(2,\R)$.
But, as mentioned in the Introduction, important differences show
up for large energy-momenta.

To understand the conjugacy classes in $\SL(2,\R)$, it is
handy to use the representation
\[
  \SL(2,\R) = \left\{ \left(
     \begin{array}{cc}
        a+b &  c+d \\
        c-d & a-b
     \end{array}
    \right)
    \; :\; a,b,c,d\in \R,\; a^2-b^2-c^2+d^2 =1 \right\}
\]
which says $\SL(2,\R)$ is geometrically a `unit hyperboloid' in a space
of signature $(+ - - +)$.  Since conjugate matrices have the same
eigenvalues, the {\em trace} and thus the number $a$ is an invariant
of conjugacy classes.  It is not a complete invariant, but it is except
for matrices with $\tr g = \pm 2$.  Every matrix in $\SL(2,\R)$ is
conjugate to one of these five kinds:
\[
\begin{array}{cccc}
 && \text{conjugate to...} & \text{trace} \\
\cline{3-4}

\text{rotations}
   & \xy (0,0)*{\includegraphics{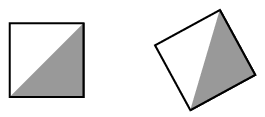}};(-3,0)*{\mapsto};\endxy
   & \rot{\alpha} & -2 \leq \tr g \leq 2 \\
\text{boosts}
   & \xy (0,0)*{\includegraphics{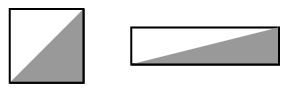}};(-3,0)*{\mapsto};\endxy
   & \twobytwo{e^\alpha}{0}{0 }{e^{-\alpha}} & \tr g \geq 2 \\
\text{antiboosts}
   & \xy (0,0)*{\includegraphics{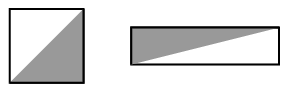}};(-3,0)*{\mapsto};\endxy
   & \twobytwo{-e^\alpha}{0}{0 }{-e^{-\alpha}} & \tr g \leq -2 \\
\text{shears}
   & \xy (0,0)*{\includegraphics{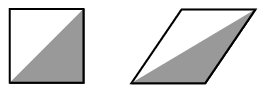}};(-3,0)*{\mapsto};\endxy
   & \twobytwo{1}{\alpha}{0}{1} & \tr g = 2 \\
\text{antishears}
   & \xy (0,0)*{\includegraphics{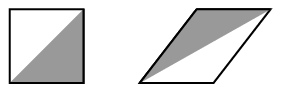}};(-3,0)*{\mapsto};\endxy
   & \twobytwo{-1}{\alpha}{0}{-1} & \tr g = -2 .
\end{array}
\]

Some explanation of this table is in order.
Every `rotation' maps to a rotation in the connected
Lorentz group $\SO_0(2,1)$: in other words, a transformation
that preserves a timelike vector in 3d Minkowski spacetime.
Similarly, every `boost' maps to
a transformation that preserves a spacelike vector, and every
`shear' maps to a transformation that preserves a lightlike vector.
Since the two-to-one map from $\SL(2,\R)$ to $\SO_0(2,1)$ maps
the matrix $-1$ to the identity, `antiboosts' get mapped to the
same elements as boosts, and `antishears' get mapped to the same
elements as shears.  (An `antirotation' would be just another rotation.)

The above chart counts certain conjugacy classes more than once.
First of all, there is an overlap at $\tr g = 2$, since
the identity rotation is also the identity shear and identity boost.
Similarly, there is an overlap at $\tr g = -2$, since
a rotation by $\pi$ is also an antishear and an antiboost.
Finally, all shears (resp.\ antishears) with $\alpha > 0$ are
conjugate to each other, and all shears (resp.\ antishears)
with $\alpha < 0$ are conjugate to each other.   These are all
the redundancies.

Knowing this, we can list all the conjugacy classes in $\SL(2,\R)$
without any redundancies.  However, it
is less tiresome to list the conjugacy classes in $\SO_0(2,1)$,
since the elements $\pm g \in \SL(2,\R)$ get identified in
$\SO_0(2,1)$, so we do not need to worry about `antiboosts'
and `antishears'.

Here are all the conjugacy classes in $\SO_0(2,1)$, and the
corresponding five types of spin-zero particles:
\begin{enumerate}
\item For any $0 < m < 2\pi/\kappa$ there is a conjugacy class containing
the image of
\[        \rot{\kappa m/2} \in \SL(2,\R).\]
This corresponds to a tardyon of mass $m$.
\item For any $0 < m < \infty$ there is a conjugacy class containing
the image of
\[ \twobytwo{e^{\kappa m/2}}{0}{0}{e^{-\kappa m/2}} \in \SL(2,\R).\]
This corresponds to a tachyon of mass $im$.
\item There is a conjugacy class containing the image of
\[  \twobytwo{1}{1}{0}{1} \in \SL(2,\R).\]
This corresponds to a positive-energy luxon.
\item There is a conjugacy class containing the image of
\[       \twobytwo{1}{-1}{0}{1} \in \SL(2,\R).\]
This corresponds to a negative-energy luxon.
\item There is a conjugacy class containing the image of
\[     \twobytwo{1}{0}{0}{1} \in \SL(2,\R). \]
This corresponds to a particle of vanishing energy-momentum.
\end{enumerate}
The factors of $1/2$ here arise from the double cover
$\SL(2,\R) \to \SO_0(2,1)$.  As explained in the Introduction,
masses of tardyons really take values in the circle
$\R/{2\pi\over \kappa}\Z$.

Each conjugacy class $Q \subseteq \SO_0(2,1)$ admits an invariant
measure which is unique up to an overall scale.  So, Theorem
\ref{3dtheorem} applies: we can form a
Hilbert space $L^2(Q)$ for particles of type $Q$, and more generally
an $n$-particle Hilbert space $L^2(Q^n)$, on which the braid group
and $\SO_0(2,1)$ gauge transformations act as unitary transformations.

{\boldmath
\section{Strings in 4d $BF$ Theory}
}
\label{4d}

All the work in the previous two sections generalizes nicely from
3 to 4 dimensions, using the loop braid group as a substitute
for the braid group.  Let space be $\R^3$ with $n$ unknotted and
unlinked circles removed:
\[
X = S - \Sigma, \qquad S = \R^3,
\qquad \Sigma = \ell_1 \cup \cdots \cup \ell_n.
\]
The fundamental group of $X$ is the free group on $n$ generators, so
for any Lie group $G$ we have
\[     \hom(\pi_1(X),G) = G^n  .\]
As explained in Section \ref{motion}, a point in this space represents
a $G$-bundle with flat connection over $X$, mod gauge transformations that
equal the identity at a chosen basepoint.   The $n$ elements of $G$
describing this point are just the holonomies around the circles
$\ell_1, \dots, \ell_n$.  Physically, we think of these circles as 
string-like `topological defects' where the flat connection on space
becomes singular.  

We explained quite generally in Section \ref{motion} how the motion group
$\Mo(S,\Sigma)$ acts on $\hom(\pi_1(X),G)$.  In the present case the
motion group is just the loop braid group $LB_n$, and its generators
act on $\hom(\pi_1(X),G) = G^n$ as follows:
\[   (g_1, \dots, g_i, g_{i+1}, \dots, g_n)s_i  =
(g_1, \dots, g_{i+1}, g_i, \dots, g_n) ,\]
\[   (g_1, \dots, g_i, g_{i+1}, \dots, g_n)\sigma_i  =
(g_1, \dots, g_{i} g_{i+1} g_{i}^{-1}, g_i, \dots, g_n)  .\]
This is easy to see using pictures.  For example, the generator
$\sigma_1$ has the following effect:
\[
\smalllabels
\xy  0;<1cm,0cm>:
(0,0)*{\includegraphics[height=4cm]{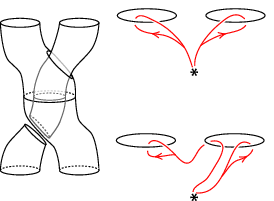}};
(.5,1.2)*{g_1};
(2,1.3)*{g_2};
(.5,-1.3)*{g_2};
(2.5,-1.2)*{g_1};
\endxy
\]
By an argument like the one we made in Section \ref{3d}
for the ordinary braid group action in 3d $BF$ theory, it
follows that $\sigma_1$ acts on the holonomies $g_1,g_2$ by switching
them while left conjugating $g_2$ by $g_1$:
\[
\def\objectstyle{\scriptstyle}
\def\labelstyle{\scriptstyle}
\xy 0;<1cm,0cm>:
(0,0)*{\xy 0;<1cm,0cm>:
      (0,0)*{\brtl};
      (-.35,1)*{g_1};
      (.35,1)*{g_2};
      (-.6,-1)*{g_1 \zap g_2};
      (.35,-1)*{g_1};
      (3,0)*{};
      (-3,0)*{};
      \endxy};
\endxy
\]
Similarly, the inverse of $\sigma_1$ acts to switch
the group elements while right conjugating $g_1$ by $g_2$:
\[
\def\objectstyle{\scriptstyle}
\def\labelstyle{\scriptstyle}
{\xy 0;<1cm,0cm>:
      (0,0)*{\brtlinv};
      (-.35,1)*{g_1};
      (.35,1)*{g_2};
      (-.35,-1)*{g_2};
      (.6,-1)*{g_1 \paz g_2};
      (3,0)*{};
      (-3,0)*{};
      \endxy}
\]
The generator $s_1$ simply switches the holonomies
$g_1$ and $g_2$:
\[
\def\objectstyle{\scriptstyle}
\def\labelstyle{\scriptstyle}
{\xy 0;<1cm,0cm>:
      (0,0)*{\brol};
      (-.35,1)*{g_1};
      (.35,1)*{g_2};
      (-.35,-1)*{g_2};
      (.35,-1)*{g_1};
      (3,0)*{};
      (-3,0)*{};
      \endxy}
\]
It is easy to see that if $G$ is unimodular, this action of
the loop braid group on $G^n$ gives rise to a unitary representation
of the loop braid group on $L^2(G^n)$.  And, just as in 3 dimensions, we can
generalize this result to the case of a quandle:

\hskip .1em

\begin{theorem} \label{4dtheorem} \et
Suppose $Q$ is a topological quandle equipped
with an invariant measure.  Then there is a unitary representation
$\rho$ of the loop braid group $LB_n$ on $L^2(Q^n)$ given by
\[         (\rho(\sigma) \psi)(q_1,\dots, q_n) =
                \psi((q_1, \dots, q_n)\sigma) \]
for all $\sigma \in LB_n$, where $LB_n$ has a right action on $Q^n$
given by:
\[
\begin{array}{ccl}
      (q_1, \dots, q_i, q_{i+1}, \dots, q_n)s_i &=&
(q_1, \dots, q_{i+1}, q_i, \dots, q_n)   \\
        (q_1, \dots, q_i, q_{i+1}, \dots, q_n)\sigma_i &=&
(q_1, \dots, q_i \zap q_{i+1}, q_i, \dots, q_n)
\end{array}
\]
There is also a unitary operator $U(q)$ on $L^2(Q^n)$ for
each element $q \in Q$, given by
\[     (U(q)\psi)(q_1, \dots, q_n)  =
    \psi(q \zap q_1, \dots, q \zap q_n)  .\]
\end{theorem}
 
\proof   While the proof is straightforward, it is worth comparing
Theorem 5.1 of Fenn, Rim\'anyi and Rourke \cite{FRR}.  This says
that the braid permutation group $BP_n$ is the group of automorphisms 
of the free quandle on $n$ generators.  Since $BP_n$ is isomorphic
to the loop braid group $LB_n$, it follows that $LB_n$ acts on $Q^n$ 
for any quandle $Q$.  The action is precisely as above.
\qed

Let us illustrate these ideas in the case where the gauge
group is the connected Lorentz group $\SO_0(3,1)$ or its
double cover $\SL(2,\C)$.  With either of these gauge groups, $BF$
theory in 4 dimensions is sometimes called `topological gravity'.

In Section \ref{quandle} we recalled the classification of 
conjugacy classes in $\SO_0(2,1)$ and its double cover 
$\SL(2,\R)$.  The classification for $\SO_0(3,1)$ and its
double cover $\SL(2,\C)$ is very similar, but simpler, because
every complex number has a square root.  It is also more 
familiar, since any element of
\[
  \SL(2,\C) = \left\{ \left(
     \begin{array}{cc}
        a &  b \\
        c &  d 
     \end{array}
    \right)
    \; :\; a,b,c,d\in \C,\; ad - bc = 1 \right\}
\]
gives a fractional linear transformation
\[     z \mapsto \frac{az + b}{cz + d}.  \]
Such transformations are precisely the
conformal transformations of the Riemann sphere.
Note that both $1$ and $-1$ in $\SL(2,\C)$
map to the identity fractional linear transformation, so the
conformal group of the Riemann sphere is 
\[        \SL(2,\C)/\{\pm 1\} \iso \SO_0(3,1) .\]
Indeed, Lorentz transformations can be thought of as 
conformal transformations of the `celestial sphere': 
the set of light rays through an observer at the origin \cite{PR}. 
A list of conjugacy classes in $\SO_0(3,1)$ can
thus be read off from the well-known classification of 
conformal transformations of the Riemann sphere \cite{Needham}.
But in fact, it is easy enough to construct this list from first
principles.

Every element of $\SO_0(3,1)$ is either conjugate to the image of the 
shear
\[          \twobytwo{1}{1}{0}{1} \in \SL(2,\C)  \]
or conjugate to the image of 
\[          \twobytwo{\lambda}{0}{0}{\lambda^{-1}}  \in \SL(2,\C) \]
for some $\lambda \ne 0$.  The conjugacy class of the latter element is
unchanged if we make the replacement $\lambda \mapsto 1/\lambda$, and
its image in $\SO_0(3,1)$ is unchanged if we make the replacement
$\lambda \mapsto -\lambda$.  These replacements (and their composite)
are the only ways we can change $\lambda$ without changing the conjugacy
class of the corresponding element of $\SO_0(3,1)$.  Using this, we can 
see there there are five types of conjugacy classes in $\SO_0(3,1)$:
\begin{enumerate}
\item For any real $m$ with 
$0 < m \le \pi/\kappa$ there is a conjugacy class containing
the image of
\[    \twobytwo{e^{i\kappa m/2}}{0}{0}{e^{-i\kappa m/2}}  \in \SL(2,\C).\]
An element conjugate to one of this form is called {\bf elliptic}.
\item For any purely imaginary $m$ with $0 < {\rm Im}(m) < \infty$ 
there is a conjugacy class containing the image of 
\[ \twobytwo{e^{i\kappa m/2}}{0}{0}{e^{-i\kappa m/2}} \in \SL(2,\C).\]
An element conjugate to one of this form is called {\bf hyperbolic}.
\item For any $m \in \C$ with $0 < {\rm Re}(m) < 2\pi/\kappa$ 
and $0 < {\rm Im}(m) < \infty$ there is a conjugacy class containing
the image of 
\[ \twobytwo{e^{i\kappa m/2}}{0}{0}{e^{-i\kappa m/2}} \in \SL(2,\C).\]
An element conjugate to one of this form is called {\bf loxodromic}.
\item There is a conjugacy class containing the image of
\[  \twobytwo{1}{1}{0}{1} \in \SL(2,\C).   \]
An element conjugate to one of this form is called {\bf parabolic}.
\item There is a conjugacy class containing the image of
\[     \twobytwo{1}{0}{0}{1} \in \SL(2,\C). \]
This class contains only the identity element. 
\end{enumerate}
 
Now let us return to $BF$ theory with gauge group $\SO_0(3,1)$, taking
space to be $\R^3$ with a collection of unknotted unlinked circles 
$\ell_1, \dots, \ell_n$ removed.  For brevity let us call these 
circles `closed strings'.   A flat connection on space will have some
holonomy $g_i \in \SO_0(3,1)$ around the $i$th string.  The above 
list of conjugacy classes lets us list possible `types' of strings,
just as we used conjugacy classes in $\SO_0(2,1)$ to list types of 
point particles in 3d gravity:

\begin{enumerate}
\item
If $g_i$ is elliptic, it acts on Minkowski spacetime as a spatial 
rotation in some reference frame.  In this reference frame, 
parallel transport around the string $\ell_i$ is a spatial rotation by 
some angle $0 < \theta \le \pi$ about some axis.
(A rotation by an angle $\theta > \pi$ is a rotation by $\theta - \pi$ 
about the opposite axis.)   This angle $\theta$ is proportional to the 
real number $m$ which appears in item 1 of the above list, as follows:
\[           \theta = \kappa m. \]
By analogy to 3d gravity, we could call the string a {\bf tardyon} in this 
case, and call the number $m$ its `mass density'.  The number $m$ is
real and takes values $0 < m \le \pi/\kappa$.

\item
If $g_i$ is hyperbolic, it acts on Minkowski spacetime as a boost in some 
reference frame.  In this reference frame, parallel transport around the 
string $\ell_i$ is a boost with rapidity $0 < \beta < \infty$ along some axis.
The rapidity $\beta$ is proportional to the imaginary 
number $m$ which appears in item 2 of the above list, as follows:
\[          \beta = \kappa {\rm Im}(m)  .\]
By analogy to 3d gravity, we could call the string a {\bf tachyon} in 
this case, and call the number $m$ its `mass density'.  The number $m$
is purely imaginary and takes values in the upper half of the imaginary 
axis: $0 < {\rm Im}(m) < \infty$.

\item 
If $g_i$ is loxodromic, it acts on Minkowski spacetime as a combined
rotation and boost about the same axis in some reference frame.  
In this reference frame, parallel transport around the string $\ell_i$
is a combination of a rotation by an angle $0 < \theta < 2\pi$ and 
a boost with rapidity $0 < \beta < \infty$ about the same axis, where
\[         \theta = \kappa {\rm Re}(m) , \qquad 
           \beta = \kappa {\rm Im}(m)  .
\]
This case has no analogue in 3d gravity.  We can still think of
$m$ as some sort of mass density, but it is complex, with 
$0 < {\rm Re}(m) < 2\pi/\kappa$ and $0 < {\rm Im}(m) < \infty$.

\item 
If $g_i$ is parabolic, it acts on Minkowski spacetime as a Lorentz
transformation fixing a single null vector.   By analogy to 3d
gravity, we could call the string a {\bf luxon} in this case, and
say $m = 0$.

\item 
If $g_i$ is the identity, we can say the string carries no 
energy-momentum, and again say $m = 0$.
\end{enumerate}

Each of these conjugacy classes $Q \subseteq \SO_0(3,1)$ is
a quandle.  The question then arises which of these quandles
admits an invariant measure, and whether this measure is unique
up to scale.  One can work this out on a case-by-case basis.

One important case is when $Q$ is the conjugacy class 
containing all rotations by some fixed angle $0 < \theta < \pi$.
This conjugacy class corresponds to a `tardyonic' closed 
string with a given mass density $0 < m \le \pi/\kappa$.  
It is easy to see that this conjugacy class $Q$
indeed admits an invariant measure.  To see this, note
that to specify a rotation by the angle $\theta$
one must first pick a future-pointing
unit timelike vector $u \in \R^4$, to split Minkowski spacetime
into space and time, and then pick a unit spacelike vector $v$ orthogonal 
to $u$, to serve as the axis of rotation.  The allowed choices of
$u$ lie in the hyperboloid
\[  H = \{ (t,x,y,z) \colon \; t^2 - x^2 - y^2 - z^2 = 1, \; t > 0 \} .\]
This hyperboloid $H$ is a Riemannian submanifold of $\R^4$.
An allowed choice of $u$ together with $v$ amounts to a point in $SH$, 
the unit sphere bundle of $H$.  So, we have $Q \iso SH$.  Since the 
unit sphere bundle of a Riemannian manifold is itself a Riemannian manifold 
in a natural way, we get a well-defined Lebesgue measure on $SH$ and thus 
$Q$, which is invariant under $\SO_0(3,1)$, since our construction
respected the Lorentz group symmetry.

Given an invariant measure on $Q$, we obtain a 
Hilbert space $L^2(Q^n)$ for $n$ strings of type $Q$.  Note that
we do not try to `symmetrize' the states in this Hilbert space.  
Instead we describe the statistics using a representation of 
the loop braid group, following Theorem \ref{4dtheorem}.  Of 
course, one should work out the details explicitly, but we leave 
this for future research.

\section{Conclusions}

Much more needs to be done to ferret out the physical significance
of the theory we have been considering here.  First, there are some
nice projects for the mathematician.  One should determine for various 
Lie groups $G$ which conjugacy classes $Q \subseteq G$ admit invariant 
measures, and when these invariant measures are unique up to an overall 
scale.  We have only done this for $G = \SO_0(2,1)$, but for applications 
to 4d physics other groups are more relevant---especially the Lorentz, 
Poincar\'e, deSitter and anti-deSitter groups.  Then, given a conjugacy 
class $Q \subseteq G$ with an invariant measure, one should work out 
explicitly the representation of the loop braid group $LB_n$ on the 
Hilbert space $L^2(Q^n)$, if possible decomposing this representation
into irreducibles, so as to understand in detail the workings of the
exotic statistics.  It would also be interesting to study how, in 
the $\kappa \to 0$ limit, the exotic statistics approach ordinary 
bosonic statistics.

For the physicist, one interesting project would be to study 
the {\it dynamics} and {\it interactions} of the `closed strings'
discussed at the purely kinematical level here.   In a paper with 
Perez \cite{BaezPerez} we describe a Lagrangian whereby these objects 
can couple to the fields in $BF$ theory.  We work out the equations 
of motion and propose a strategy for quantizing the resulting theory, 
analogous to the known quantization of point particles coupled to 3d 
gravity \cite{NouiPerez}.   

A more ambitious project would be to generalize all our results from 
collections of unlinked unknotted circles to arbitrary embedded graphs.
Finally, a still more ambitious project would be to use these ideas as part 
of a perturbative expansion of MacDowell--Mansouri gravity about 4d $BF$
theory, as proposed by Freidel and Starodubtsev \cite{FreidelStarodubtsev}.

\section{Appendix}

Here we present a proof of Theorem \ref{presentation2} on 
p.\ \pageref{presentation2}.

\vskip 1em

\proof We begin by demonstrating that the relations in the
statement of Theorem \ref{presentation2} follow from those given
in Theorem \ref{presentation1}.  It clearly suffices to show that
the relations in $(b^{\prime})$ and $(c^{\prime})$ follow from the
relations in $(a), (b)$ and $(c)$.

In what follows, we make frequent use of the correspondence
between generators $\sigma_{ij}$ of $PLB_n$ and generators
$\sigma_i$ of $LB_n$ as given in (\ref{generatorCorrespondence1})
and (\ref{generatorCorrespondence2}).  In fact, since these follow
from different relations in the presentation of Theorem
\ref{presentation1}, it suffices for our purposes to take one
expression from each of these, say
\begin{equation}
\label{generatorCorrespondence}
   \sigma_{ij}
= \left\{
\begin{array}{cl}
s_i s_{i+1} \cdots s_{j-1} \sigma_{j-1} s_{j-2} \cdots
s_{i+1}s_i & \textrm{for \;} i < j \\
s_j s_{j+1} \cdots s_{i-2} \sigma_{i-1} s_{i-1} \cdots s_{j+1} s_j
& \textrm{for \;} i > j
\end{array}
\right.
\end{equation}
These representations of $\sigma_{ij}$ follow directly from the
definition of $\sigma_i$ along with the relations (\ref{rel7}),
(\ref{rel9}), and (\ref{rel11}).
\\

$\bullet$ Relation (\ref{Rel7}):  We wish to show that $s_j
\sigma_i = \sigma_i s_j$ for $|i-j| > 1.$ To check this, we begin
with relation (\ref{rel8}) in the form:
$$s_j \sigma_{i(i+1)}
= \sigma_{i(i+1)}s_j,$$  where $|i - j|
>1.$  Using (\ref{generatorCorrespondence}) above, this becomes:
$$s_j s_i \sigma_i = s_i \sigma_i s_j.$$
Applying relation (\ref{rel1}) of to the left-hand side and then
cancelling $s_i$ from each side gives $s_j \sigma_i = \sigma_i
s_j$ when $|i - j| > 1,$ which is (\ref{Rel7}). \\

$\bullet$ Relation (\ref{Rel8}):  We wish to show that $s_i
s_{i+1} \sigma_i = \sigma_{i+1} s_i s_{i+1}$ for $1 \le i \le
n-2.$  Beginning with relation (\ref{rel9}) with $j = i+1$, we
obtain:
$$s_{i+1} \sigma_{i(i+1)} = \sigma_{i(i+2)} s_{i+1}.$$
By (\ref{generatorCorrespondence}) this gives:
$$s_{i+1} s_i \sigma_i = s_i s_{i+1} \sigma_{i+1} s_i s_{i+1}.$$

\noindent Multiplying on the right by $s_{i+1} s_i$ and on the
left by $s_{i} s_{i+1}$, we have:
\begin{align*}
\sigma_i s_{i+1} s_i &= s_i s_{i+1} s_i s_{i+1} \sigma_{i+1}  && \\
                     &=  s_i s_i s_{i+1} s_i \sigma_{i+1}  && \textrm{by
(\ref{rel2})} \\
                     &= s_{i+1} s_i \sigma_{i+1}  &&  \textrm{by
                     (\ref{rel3})}
\end{align*}
This can be rewritten as $s_i s_{i+1} \sigma_{i} = \sigma_{i+1}s_i
s_{i+1},$ which is (\ref{Rel8}).
\\

$\bullet$ Relation (\ref{Rel9}): We wish to show that $\sigma_i
\sigma_{i+1} s_i = s_{i+1} \sigma_i \sigma_{i+1}$ for $1 \le i \le
n-2.$ To verify this we use relation (\ref{rel5}) with $i, i+1$
and $i+2$, which gives:
$$\sigma_{i(i+2)}\sigma_{(i+1)(i+2)} = \sigma_{(i+1)(i+2)}\sigma_{i(i+2)}.$$

\noindent By (\ref{generatorCorrespondence}) this becomes:
$$(s_i s_{i+1} \sigma_{i+1} s_{i})(s_{i+1} \sigma_{i+1})  = (s_{i+1}
\sigma_{i+1})
(s_i s_{i+1} \sigma_{i+1} s_{i}).$$

\noindent Applying relation (\ref{Rel8}) on the left hand side
gives:
$$s_i s_{i+1} s_i s_{i+1} \sigma_{i} \sigma_{i+1}  = (s_{i+1} \sigma_{i+1})
(s_i s_{i+1} \sigma_{i+1} s_{i}).$$

\noindent Multiplying by $s_i s_{i+1} s_i$ on the left produces:
\begin{align*}
s_{i+1} \sigma_{i} \sigma_{i+1} &= s_i s_{i+1} s_i s_{i+1}
\sigma_{i+1} s_i s_{i+1} \sigma_{i+1} s_i && \\
& = s_{i+1} s_i \sigma_{i+1} s_i s_{i+1} \sigma_{i+1} s_i &&
\textrm{by (\ref{Rel2})} \\
& = \sigma_{i} \sigma_{i+1} s_i && \textrm{by (\ref{Rel8})}
\end{align*}
which is (\ref{Rel9}). \\

$\bullet$ Relation (\ref{Rel4}):  We wish to show that $\sigma_i
\sigma_j = \sigma_j \sigma_i$ for $|i-j| > 1$. To do so, we use
relation (\ref{rel4}) with $i, i+1, j, j+1$, which are clearly all
distinct for $|i-j|>1$.  We therefore have:
$$\sigma_{i(i+1)} \sigma_{j(j+1)} = \sigma_{j(j+1)}
\sigma_{i(i+1)},$$ which, by (\ref{generatorCorrespondence}),
becomes: $$s_i \sigma_i s_j \sigma_j = s_j \sigma_j s_i \sigma_i.
$$  Applying (\ref{Rel7}) to both sides of this
equation, followed by relation (\ref{rel1}), we obtain:
$$\sigma_i \sigma_j = \sigma_j \sigma_i$$ with $|i - j|
>1,$ which is (\ref{Rel4}). \\

$\bullet$ Relation (\ref{Rel5}):  We wish to show that $\sigma_i
\sigma_{i+1} \sigma_i = \sigma_{i+1} \sigma_i \sigma_{i+1}$ for $1
\le i \le n-2$. To check this we start with relation (\ref{rel6})
with $i, i+1,$ and $i+2$, which are clearly all distinct. Thus, we
have:
$$\sigma_{i(i+1)} \sigma_{(i+2)(i+1)} \sigma_{i(i+2)} = \sigma_{i(i+2)}
\sigma_{(i+2)(i+1)} \sigma_{i(i+1)}.$$
\noindent Using the correspondence given in
(\ref{generatorCorrespondence}) and cancelling $s_i$ from both
sides, we obtain:
\begin{align*}
s_{i+1} \sigma_{i+1} s_i \sigma_{i+1} s_{i+1} s_i \sigma_i &=
\sigma_i \sigma_{i+1} s_{i+1} s_i s_{i+1} \sigma_{i+1} s_i && \\
&= \sigma_i \sigma_{i+1} s_{i} s_{i+1} s_{i} \sigma_{i+1} s_i
 && \textrm{by (\ref{rel2})} \\
&= \sigma_i \sigma_{i+1} s_{i} \sigma_i s_{i+1} && \textrm{by
(\ref{Rel8}), (\ref{rel3})} \\
&= s_{i+1} \sigma_i \sigma_{i+1} s_{i+1} && \textrm{by
(\ref{Rel9})}.
\end{align*}

\noindent Cancelling $s_{i+1}$ on the left and multiplying by
$s_{i+1}$ on the right produces:
\begin{align*}
\sigma_i \sigma_{i+1} \sigma_i  &= \sigma_{i+1} s_i \sigma_{i+1}
s_{i+1} s_i \sigma_i s_{i+1} && \\
& = \sigma_{i+1} \sigma_i \sigma_{i+1} &&
\end{align*}
where in the last step we used (\ref{rel5}) in the form $s_i
\sigma_i \sigma_{i+1} s_{i+1} =
\sigma_{i+1} s_{i+1} s_i \sigma_i$. This is (\ref{Rel5}). \\

The loop braid group thus has generators that satisfy all of the
relations of the braid permutation group.  It remains to show that
these relations are sufficient, which we do by demonstrating that
the relations in the statement of Theorem \ref{presentation1}
follow from those given in Theorem \ref{presentation2}.  In this 
direction of the proof it is convenient to use both of the equivalent
expressions (\ref{generatorCorrespondence1}) and 
(\ref{generatorCorrespondence2}) as the correspondence between
generators $\sigma_i$ and $\sigma_{ij}$.\\


$\bullet$ Relation (\ref{rel7}):  This relation simply says 
$s_i \sigma_{i(i+1)} = \sigma_{(i+1)i}s_i$, which is immediate from (\ref{generatorCorrespondence})
since both sides are equal to $\sigma_i$. \\

$\bullet$ Relation (\ref{rel8}):  We wish to show $s_k \sigma_{ij}=\sigma_{ij}s_k$, 
whenever $i,j,k,k+1$ are distinct.  
When either $k+1<i<j$ or $i<j<k$, $s_k$ commutes with each of the factors in the expansion
\[
   \sigma_{ij}
=
s_i s_{i+1} \cdots s_{j-1} \sigma_{j-1} s_{j-2} \cdots
s_{i+1}s_i 
\]
by (\ref{Rel1}) and (\ref{Rel7}).  
Similarly, when $k+1<j<i$ or $j<i<k$, $s_k$ commutes with each factor in
\[
\sigma_{ij}=s_j s_{j+1} \cdots s_{i-2} \sigma_{i-1} s_{i-1} \cdots s_{j+1} s_j.
\]
When $i<k<k+1<j$ we also need two 
applications of (\ref{Rel2}):
\begin{align*}
s_k \sigma_{ij} &=  s_k s_i \cdots s_{j-1}\sigma_{j-1} s_{j-2} \cdots s_i \\
    &= s_i \cdots s_{k-2}s_k s_{k-1} s_k s_{k+1} \cdots s_{j-1} \sigma_{j-1} s_{j-2} \cdots s_i 
       && \textrm{by (\ref{Rel1})} \\
    &= s_i \cdots s_{k-2}s_{k-1} s_{k} s_{k-1} s_{k+1} \cdots s_{j-1} \sigma_{j-1} s_{j-2} \cdots s_i 
       && \textrm{by (\ref{Rel2})} \\
    &= s_i \cdots s_{k-2}s_{k-1} s_{k} s_{k+1} \cdots s_{k-1}s_{j-1} \sigma_{j-1} s_{j-2} \cdots s_i
       && \textrm{by (\ref{Rel1})} \\
    &= s_i \cdots s_{j-1} \sigma_{j-1} s_{j-2} \cdots s_{k+1}s_{k-1}s_k s_{k-1} s_{k-2} \cdots  s_i 
       && \textrm{by (\ref{Rel1}), (\ref{Rel7})} \\
    &= s_i \cdots s_{j-1} \sigma_{j-1} s_{j-2} \cdots s_{k+1}s_{k}s_{k-1} s_{k} s_{k-2} \cdots  s_i 
       && \textrm{by (\ref{Rel2})} \\
    &= \sigma_{ij} s_k
       && \textrm{by (\ref{Rel1})} 
\end{align*}
The only remaining case is $j<k<k+1<i$, which is handled similarly. \\

$\bullet$ Relation (\ref{rel9}):  We wish to show that $s_j \sigma_{ij} = \sigma_{i(j+1)}s_j$ whenever $i \neq j+1$.  When $i<j$ we have:
\begin{align*}
s_j\sigma_{ij} 
                         &= s_i s_{i+1} \cdots s_j s_{j-1} \sigma_{j-1}s_{j-2}\cdots s_{i+1}s_i    
                                      && \textrm{by (\ref{Rel1})} \\
                         &= s_i s_{i+1} \cdots s_{j-1} s_j s_{j-1} s_j \sigma_{j-1}s_{j-2}\cdots s_{i+1}s_i    
                                      && \textrm{by (\ref{Rel2})} \\
                         &= s_i s_{i+1} \cdots s_{j-1} s_j \sigma_{j}s_{j-1} s_j s_{j-2}\cdots s_{i+1}s_i    
                                      && \textrm{by (\ref{Rel7})} \\
                         &= \sigma_{i(j+1)}s_j
                                      && \textrm{by (\ref{Rel1}), (\ref{generatorCorrespondence})}                                     
\end{align*}
and the case $i>j+1$ is similar. \\

$\bullet$ Relation (\ref{rel11}):  The proof that $s_i \sigma_{ij}= \sigma_{(i+1)j} s_i$ 
is essentially the same as the proof of (\ref{rel9}) above. \\

$\bullet$ Relation (\ref{rel4}):  We wish to show $\sigma_{ij}
\sigma_{k \ell} = \sigma_{k \ell} \sigma_{ij}$, whenever $i, j,
k,$ and $\ell$ are distinct.  Naively there are $4!$ orderings of $i,j,k,\ell$
to consider,
but symmetry of the relation implies only 8 are independent.
All cases are proved similarly; we demonstrate only 
the case $i < j < k < \ell$:
\begin{align*} \sigma_{ij} \sigma_{k \ell} &= (s_i
  \cdots s_{j-1} \sigma_{j-1} s_{j-2} \cdots s_i) (s_k
   \cdots s_{\ell-1} \sigma_{\ell -1} s_{\ell-2}
  \cdots s_k ) && \\
&= s_k \cdots s_{\ell-1} (s_i  \cdots
  s_{j-1} \sigma_{j-1} s_{j-2} \cdots s_i)
  (\sigma_{\ell-1} s_{\ell-2} \cdots s_k ) \;\; \textrm{by
  (\ref{Rel1}), (\ref{Rel7})} && \\
&= s_k \cdots s_{\ell-1} \sigma_{\ell - 1} (s_i
   \cdots s_{j-1} \sigma_{j-1} s_{j-2} \cdots
  s_i) (s_{\ell-2} \cdots s_k ) \; \; \textrm{by
  (\ref{Rel7}), (\ref{Rel4})} && \\
&= (s_k \cdots s_{\ell-1} \sigma_{\ell - 1}
  s_{\ell-2} \cdots s_k)(s_i  \cdots s_{j-1} 
  \sigma_{j-1} s_{j-2} \cdots s_i) \; \; \textrm{by
  (\ref{Rel1}), (\ref{Rel7})} && \\
& = \sigma_{k \ell} \sigma_{ij}. &&
\end{align*}

$\bullet$ Relation (\ref{rel5}):   We wish to show that
$\sigma_{ik} \sigma_{jk} = \sigma_{jk} \sigma_{ik}$ when $i, j,k$
are distinct.  We have three independent cases:  $i < j < k$, $i < k < j$, 
and $k < i < j$. In the case $i<j<k$, we first note that
if $j\neq i+1$, then by (\ref{rel8}) and (\ref{rel9}) we
have:
\begin{align*}
  \sigma_{ik} \sigma_{jk} &= s_{j-1}(\sigma_{ik}\sigma_{(j-1)k})s_{j-1} \\
\text{and }\;  \sigma_{jk} \sigma_{ik} &= s_{j-1}(\sigma_{(j-1)k}
\sigma_{ik})s_{j-1}.
\end{align*}
By repeated application of these facts, it suffices to consider
the subcase where $j=i+1$.  Similarly, if $k\neq j+1$, we can use
(\ref{rel8}) and (\ref{rel11})  to reduce to the case
where $k=j+1$.  Thus it suffices to consider only the cases where
$i,j,k$ are consecutive:
\begin{align*}
\sigma_{i(i+2)} \sigma_{(i+1)(i+2)} & = (s_i s_{i+1} \sigma_{i+1}
s_i)(s_{i+1} \sigma_{i+1}) && \\
& = s_i s_{i+1} s_i s_{i+1} \sigma_i \sigma_{i+1} && \textrm{by
(\ref{Rel8}) } \\
& = s_{i+1} s_i \sigma_i \sigma_{i+1} && \textrm{by (\ref{Rel2})} \\
& = s_{i+1} s_i s_{i+1} \sigma_i \sigma_{i+1} s_i &&  \textrm{by
(\ref{Rel9})} \\
& = s_{i+1} \sigma_{i+1} s_i s_{i+1} \sigma_{i+1} s_i &&
\textrm{by (\ref{Rel8})} \\
& = \sigma_{(i+1)(i+2)} \sigma_{i(i+2)}. &&
\end{align*}
This proves the case $i<j<k$.  The remaining two cases are
similar. \\

$\bullet$ Relation (\ref{rel6}):  We wish to show that
$\sigma_{ij} \sigma_{kj} \sigma_{ik} =
\sigma_{ik}\sigma_{kj}\sigma_{ij}$ when $i,j,k$ are distinct.  In
light of (\ref{rel5}) this equation is symmetric under the
interchange of $i$ and $k$, and this symmetry reduces the number
of independent cases to 3: $i<j<k$, $i<k<j$, and $j<i<k$.  In the
case $i<j<k$, we first note that if $j\neq i+1$, then by
(\ref{rel8}) and (\ref{rel9}) we have
\begin{align*}
  \sigma_{ij}\sigma_{kj} \sigma_{ik} &= s_{j-1}(\sigma_{i(j-1)}\sigma_{k(j-1)}
\sigma_{ik})s_{j-1} \\
\text{and }\;  \sigma_{ik}\sigma_{kj} \sigma_{ij} &=
s_{j-1}(\sigma_{ik}\sigma_{k(j-1)} \sigma_{i(j-1)})s_{j-1}
\end{align*}
By repeated application of these facts, it suffices to consider
the subcase where $j=i+1$.  Similarly, if $k\neq j+1$, we can use
(\ref{rel8}) and (\ref{rel11}) to reduce to the case where
$k=j+1$.  Thus it suffices to consider only the cases where
$i,j,k$ are consecutive:
\begin{align*}
\sigma_{i(i+1)}\sigma_{(i+2)(i+1)}\sigma_{i(i+2)}
    &= (s_i \sigma_i)( \sigma_{i+1} s_{i+1}) (s_i s_{i+1} \sigma_{i+1} s_i) \\
    &= s_i \sigma_i \sigma_{i+1} s_{i} s_{i+1} s_{i} \sigma_{i+1} s_i &&
\text{by (\ref{Rel2})} \\
    &= s_i \sigma_i \sigma_{i+1} s_{i} \sigma_i s_{i+1}      && \text{by
(\ref{Rel8})} \\
    &= s_i s_{i+1} \sigma_i \sigma_{i+1} \sigma_i s_{i+1}      && \text{by
(\ref{Rel9})} \\
    &= s_i s_{i+1} \sigma_{i+1} \sigma_{i} \sigma_{i+1} s_{i+1}      && \text{by
(\ref{Rel5})} \\
&=\sigma_{i(i+2)}  \sigma_{i(i+1)} \sigma_{(i+2)(i+1)} \\
&=\sigma_{i(i+2)} \sigma_{(i+2)(i+1)} \sigma_{i(i+1)}   
        && \text{by (\ref{rel5})} 
\end{align*}
This proves the case of $i<j<k$.  The other two independent cases
are similar.

Thus, the relations of Theorem \ref{presentation2} imply those of
Theorem \ref{presentation1}. \qed

As pointed out by Blake Winter, one can also prove 
Theorem \ref{presentation2} as follows.  Fenn, Rim\'anyi, 
and Rourke \cite{FRR} show that the braid permutation group 
$BP_n$ is isomorphic to the subgroup, of $\Aut(F_n)$ generated 
by all permutations of basis elements, together with all operations 
of conjugating one basis element by another.  Let $X$ be $\R^3$
with unlinked unknotted circles $\ell_1, \dots, \ell_n$ removed.
As we have seen, $\pi_1(X) = F_n$, the free group on $n$ generators,
so by the work of Dahm, the loop braid group acts as automorphisms
of $F_n$.  Let $D \maps LB_n \to \Aut(F_n)$ be the resulting homomorphism.
Goldsmith \cite{Goldsmith} shows that the image of $D$ is precisely
the above subgroup of $\Aut(F_n)$ and that, moreover, $D$ is one-to-one.  
It follows that $LB_n$ and $BP_n$ are isomorphic.  Since Fenn, Rim\'anyi
and Rourke prove that $BP_n$ has the presentation given in Theorem 
\ref{presentation2}, it follows that $LB_n$ also has this presentation.

After this paper appeared on the arXiv, Sumati Surya pointed out
that Theorem \ref{presentation2} can also be proved using
results of Fuks-Rabinowitz \cite{FR} and McCullough and 
Miller \cite{McCulloughM}.

\subsubsection*{Acknowledgements}

Most of all we would like to thank Xiao-Song Lin for allowing us
to read a draft of his paper on the loop braid group.  This
inspired us to write the present paper.  We would also like to
thank A.\ P.\ Balachandran, James Dolan, Sam Nelson, Carlo Rovelli, 
Sumati Surya, and Blake Winter.


\begin{thebibliography}{10}

\bibitem{ABKS} C.\ Aneziris, A.\ P.\ Balachandran, L.\ 
Kauffman and A.\ M.\ Srivastava, Novel statistics for strings
and string `Chern--Simons' terms, {\sl Int.\ J.\ Mod.\ Phys.\ }
{\bf A6} (1990), 2519--2558.  
 
\bibitem{Baez} J.\ C.\ Baez,
An introduction to spin foam models of BF theory and quantum
gravity, in {\sl Geometry and Quantum Physics}, eds.\
H.\ Gausterer and H.\ Grosse, Lecture Notes in Physics {\bf 543},
Springer, Berlin, 2000, pp. 25-93.  Also available as
gr-qc/9905087.

\bibitem{BaezPerez} J.\ C.\ Baez and A.\ Perez,  
Quantization of strings and branes coupled to $BF$ theory,
to appear.

\bibitem{Balachandran} 
A.\ P.\ Balachandran, Statistics, strings and gravity, in
{\sl Mathematical Physics: Proceedings, 3rd Regional
Conference, Islamabad, Pakistan, February 17--24, 1989}, 
eds.\ F. Hussain and A.\ Qadir, World Scientific, Singapore, 1990,
pp.\ 31-45. 


\bibitem{BKS} P.\ Bonderson, A.\ Kitaev, and K.\ Shtengel,
Detecting non-abelian statistics in the $\nu = 5/2$
fractional quantum Hall state, {\sl Phys.\ Rev.\ Lett.\ 96} (2006)
016803.  Also available as cond-mat/0508616.

\bibitem{Brieskorn} E.\ Brieskorn, Automorphic sets and singularities,
in {\sl Braids}, {\sl Contemp.\ Math.\ } {\bf 78}, Amer.\ Math.\ Soc.,
Providence, RI, 1998, pp.\ 45--115.



\bibitem{CZG} F.\ E.\ Camino, W.\ Zhou and V.\ J.\ Goldman,
Realization of a Laughlin quasiparticle interferometer: observation
of fractional statistics, {\bf Phys.\ Rev.\ B} {\bf 72} (2005),
075342.  Also available as cond-mat/0502406.

\bibitem{CS} J.\ S.\ Carter and M.\ Saito,
{\sl Knotted Surfaces and their Diagrams},
Amer.\ Math.\ Soc., Providence, RI, 1998.

J.\ S.\ Carter, S.\ Kamada, and M.\ Saito,
{\sl Surfaces in 4-Space}, Springer, Berlin, 2004.

\bibitem{Dahm} D.\ M.\ Dahm, {\sl A Generalisation of Braid Theory},
Ph.D. Thesis, Princeton University, 1962.


\bibitem{FRR} R.\ Fenn, R.\ Rim\'anyi and C.\ Rourke,
The braid-permutation group, {\sl Topology} {\bf 36} (1997), 123--135.

\bibitem{FR} R.\ Fenn, C. \ Rourke, Racks and links in
codimension two, {\sl Journal of Knot Theory and its Ramifications}
{\bf 1} (1992), 343--406.

\bibitem{FKLW}  M.\ H.\ Freedman, A.\ Kitaev, M.\ J.\ Larsen and
Z.\ Wang, Topological quantum computation.  Also available as
quant-ph/0101025.

\bibitem{Freidel} L.\ Freidel and D.\ Louapre,
Ponzano-Regge model revisited I: gauge fixing, observables and
interacting spinning particles, {\sl Class.\ Quant.\ Grav.\ } {\bf 21}
(2004), 5685-5726. Also available as hep-th/0401076.

L.\  Freidel and D.\ Louapre, Ponzano-Regge model revisited II:
equivalence with Chern-Simons, available as gr-qc/0410141

L.\ Freidel and E.\ R.\ Livine, Ponzano-Regge model revisited III:
Feynman diagrams and effective field theory, available as hep-th/0502106.

\bibitem{FKS} L.\ Freidel, J.\ Kowalski-Glikman and L.\ Smolin,
2+1 gravity and doubly special relativity,
{\sl Phys.\ Rev.\ }{\bf D69} (2004), 044001.   Also available as
hep-th/0307085.

\bibitem{FreidelStarodubtsev} L.\ Freidel and A.\ Starodubtsev,
Quantum gravity in terms of topological observables, available as
hep-th/0501191.

\bibitem{Goldsmith} D.\ L.\ Goldsmith, The theory of
motion groups.  {\sl Michigan Math.\ J.\ }{\bf 28} (1981), 3--17.


\bibitem{Goldsmith2} D.\ L.\ Goldsmith, Motion of links in the $3$-sphere,
{\sl Math.\ Scand.\ }{\bf 50} (1982), 167--205.


\bibitem{Fuks} D.\ I.\ Fuks--Rabinovich, On the automorphism groups of
free products, I (Russian), {\sl Mat.\ Sbornik} {\bf 8} (1940), 265--276.

D.\ I.\ Fuks--Rabinovich, On the automorphism groups of free products,
II (Russian), {\sl Mat.\ Sbornik} {\bf 9} {\bf 9} (1941), 183--220. 

\bibitem{J} D.\ Joyce, A classifying invariant of knots, the knot quandle,
{\sl J.\  Pure Appl.\ Alg.} {\bf 23}  (1982), 37--65.

\bibitem{K} L.\ H.\ Kauffman, {\sl Knots and Physics}, World Scientific,
Singapore, 1991.

\bibitem{Kitaev} A.\ Kitaev, Fault-tolerant quantum computation by anyons,
{\sl Ann.\ Phys.\ }{\bf 303} (2003), 2--30.  Also available as
quant-ph/9707021.

\bibitem{Kobayashi} S.\ Kobayashi,
Chap.\ 1 \S 2: Flat bundles and flat connections,
{\sl Differential Geometry of Complex Vector Bundles},
Princeton U.\ Press, Princeton, New Jersey, 1987, pp.\ 4--7.

\bibitem{Krasnov} K.\ Krasnov, Quantum gravity with matter via group
field theory, available as hep-th/0505174.

\bibitem{Lin} X.-S.\ Lin, The motion group of the unlink and its
representations, preprint, 2005.

\bibitem{MM} S.\ W.\ MacDowell and F.\ Mansouri, Unified geometric
theory of gravity and supergravity, {\sl Phys.\ Rev.\ Lett.\ }{\bf 38}
(1977), 739--742.  Erratum, {\sl ibid.} {\bf 38} (1977), 1376.

\bibitem{McCool} L.\ McCool, On basis-conjugating automorphisms of
free groups, {\sl Canad.\ J.\ Math.\ }{\bf 38} (1986), 1525--1529.


\bibitem{McCulloughM} D.\ McCullough and A.\ Miller, 
{\sl Homeomorphisms of 3-Manifolds with Compressible Boundary},
Chapter 5, {\sl Memoirs AMS} {\bf 344}, American Mathematical
Society, Providence, 1986.

\bibitem{Needham} T.\ Needham, {\sl Visual Complex Analysis}, Clarendon
Press, Oxford, 1997.

\bibitem{Niemi} A.\ Niemi, The exotic statistics of leapfrogging
smoke rings, {\sl Phys.\ Rev.\ Lett.\ }{\bf 94} (2005), 124502.
Also available as cond-mat/0410212.

\bibitem{Oriti} D.\ Oriti, {\sl Spin Foam Models of Quantum Spacetime},
Ph.D.\ thesis, University of Cambridge.  Also available as
gr-qc/0311066.

\bibitem{PR} Roger Penrose and Wolfgang Rindler, {\sl Spinors and  
Space-Time}, Vol.\ 1, Cambridge U.\ Press, Cambridge, 1985.

\bibitem{Perez} A.\ Perez, Spin foam models for quantum gravity,
{\sl Class.\ Quant.\ Grav.\ }{\bf 20} (2003), R43--R104.

\bibitem{NouiPerez} K.\ Noui and A.\ Perez, 
Three dimensional loop quantum gravity: coupling to point particles,
{\sl Class.\ Quant.\ Grav.\ }{\bf 22} (2005), 4489--4514

\bibitem{Plebanski} J.\ Plebanski, On the separation of Einsteinian
substructures, {\sl Jour.\ Math.\ Phys.\ }{\bf 18} (1977) 2511-2520.

\bibitem{Rubinsztein} R.\ L.\ Rubinsztein, On the group of motions of
oriented, unlinked and unknotted circles in $\R^3$, I.  Preprint,
Uppsala University, 2002.


\bibitem{Rubinsztein2} R.\ L.\ Rubinsztein, Topological quandles
and invariants of links, available as math.GT/0508536.

\bibitem{Surya} S.\ Surya, Cyclic statistics in three dimensions,
{\sl J.\ Math.\ Phys.\ } {\bf 45} (2004), 2515--2525.
Also available as hep-th/0308011.


\bibitem{Szabo} R.\ J.\ Szabo, Topological field theory and quantum holonomy
representations of motion groups, {\sl Ann.\ Phys.\ }{\bf 280}
(2000), 163--208.  Also available as hep-th/9908051.

\bibitem{Vershinin}  V.\ V.\ Vershinin, On homology of virtual braids
and Burau representation, available as math.GT/9904089.

\bibitem{Wattenberg} F.\ Wattenberg, Differentiable motions of unknotted,
unlinked circles in $3$-space, {\sl Math.\ Scand.\ }{\bf 30}
(1972), 107--135.


\end{thebibliography}
\end{document}